\documentclass[11pt]{article}

\usepackage[letterpaper,margin=1in]{geometry}
\usepackage[T1]{fontenc}
\usepackage[utf8]{inputenc}
\usepackage{lmodern}
\usepackage{microtype}
\usepackage{amsmath,amssymb}
\usepackage{graphicx}
\usepackage{booktabs}
\usepackage{array}
\usepackage{multirow}
\usepackage{longtable}
\usepackage[numbers,sort&compress]{natbib}
\usepackage{url}
\usepackage[hidelinks]{hyperref}
\usepackage{authblk}

\title{Hybrid quantum-classical attention for histopathology-based molecular profiling in data-limited cancers}

\author[1]{Kahn Rhrissorrakrai\thanks{Corresponding author: \href{mailto:krhriss@us.ibm.com}{krhriss@us.ibm.com}}}
\author[1]{Aritra Bose}
\author[1]{Aldo Guzm\'an-S\'aenz}
\author[1]{Filippo Utro}
\author[1]{Laxmi Parida}
\affil[1]{IBM Research, IBM T.J. Watson Research Center, 1101 Kitchawan Road, Yorktown Heights, New York 10598, USA}
\date{}

\begin{document}
\maketitle

\begin{abstract}
Molecular profiling from routine histopathology could expand access to precision oncology when sequencing is unavailable, tissue is limited, or training cohorts are small. Here, we developed a hybrid quantum-classical attention strategy for digital pathology by integrating a quantum-derived doubly stochastic attention module into a transformer model for histopathology-based gene expression prediction. Across 29 cancer cohorts from The Cancer Genome Atlas and an independent pancreatic cancer cohort from the Clinical Proteomic Tumor Analysis Consortium, the hybrid module produced selective gains relative to standard softmax attention, with the largest relative improvements in smaller and data-limited cancers, including adrenocortical carcinoma and uveal melanoma. These gains were not transcriptome-wide; instead, the model redistributed predictive performance across genes and pathways, improving biologically relevant targets in some tumor contexts while worsening others. In adrenocortical carcinoma, the molecular targets preferentially improved under hybrid attention were also enriched for genes associated with worse overall survival, linking improved molecular inference to clinically meaningful prognostic information in a rare cancer. In pancreatic cancer transfer experiments, the hybrid model improved prediction for selected metabolic and lineage-associated genes, but did not uniformly improve performance under cross-cohort shift. A leave-one-cancer-out mixed-effects analysis showed that part of the gene-level benefit was predictable from baseline molecular features, while residual analysis identified cancer-specific biological programs that improved more or less than expected. Separate quantum hardware experiments demonstrated recovery of the doubly stochastic matrix primitive underlying the attention mechanism on IBM quantum processors. Together, these findings position hybrid quantum-classical attention as a promising digital pathology strategy for molecular triage and image-based molecular profiling when direct molecular testing is unavailable, incomplete, or impractical.
\end{abstract}


\section{Introduction}
Molecular profiling informs cancer diagnosis, prognosis, treatment selection, and eligibility for targeted therapies~\cite{zhang_molecular_2020,akhoundova_clinical_2022}, but access to transcriptomic assays remains uneven across clinical settings because sequencing can be costly, tissue-consuming, and difficult to obtain for small biopsies, retrospective cohorts, and sequencing unavailable / tissue-limited settings such as rare cancers. In contrast, hematoxylin and eosin (H\&E) whole-slide images (WSIs) are routinely generated as part of standard pathology workflows~\cite{couture_deep_2022}, making digital histopathology an attractive substrate for scalable molecular inference when direct sequencing is unavailable. Recent advances in digital medicine have therefore increasingly focused on clinically relevant artificial intelligence models that extract molecular and prognostic information from routinely collected data, including pathology images~\cite{wang_predicting_2021}.

Whole-slide image models have shown that histopathology can support prediction of clinically meaningful molecular phenotypes, including biomarker status, patient risk, and gene-expression-related programs. Recent studies published in digital pathology and related translational AI venues have demonstrated the feasibility of predicting HER2 status from gastric cancer histology, integrating pathology with multimodal clinical data for genetic prescreening, and linking whole-slide image representations with molecular measurements such as spatial transcriptomics and transcriptome-wide gene expression~\cite{liao_artificial_2025, wu_deep_2025}. Together, these studies suggest that image-based molecular profiling may complement conventional assays, particularly when tissue or sequencing access is limited, but they also highlight persistent challenges in generalizability, target specificity, and performance in small cohorts.

Transformer architectures provide a flexible framework for learning from high-dimensional biological and imaging data because self-attention can model long-range dependencies among input tokens~\cite{vaswani2017attention}. In computational biology and digital pathology, attention-based models have been applied to protein structure prediction, single-cell omics analysis, histopathology image analysis and image-based molecular prediction~\cite{abramson2024accurate,yang2022scbert,chen2024towards,pizurica2024digital,zhang2024inferring,li2024spadit}. For WSI-based molecular profiling, SEQUOIA established a transformer framework that combines tile-level image feature extraction, feature aggregation and sequence modeling to regress transcriptome-wide expression from cancer histopathology images~\cite{pizurica2024digital}. Its performance depended strongly on both the visual encoder and attention mechanism, with pathology-specific representations and efficient attention improving prediction relative to more generic image features and standard attention~\cite{pizurica2024digital,chen2024towards}.

Despite these advances, image-based molecular profiling from WSIs remains challenging. Performance varies across cancer types and genes, reflecting differences in cohort size, tumor biology, tissue heterogeneity, image quality, transcriptomic variability and the degree to which molecular programs are morphologically encoded~\cite{pizurica2024digital,wang_predicting_2021}. These challenges are particularly pronounced in rare or data-limited cancers, where large training cohorts are difficult to assemble but molecular characterization may be especially valuable~\cite{chen_rare_2026,obeid_advancing_2024}. Because effective deep learning in histopathology often requires large annotated datasets and can be sensitive to domain variation, there is a need for modeling strategies that impose useful inductive biases without requiring wholesale changes to established architectures~\cite{obeid_advancing_2024,komura_machine_2025}.

One such bias is doubly stochastic attention. Standard transformer attention applies row-wise softmax normalization to token-token similarity scores, producing a right-stochastic attention matrix whose rows sum to one. This normalization does not constrain column sums and can allow imbalanced aggregate attention across tokens. Doubly stochastic matrices (DSMs), in which both rows and columns sum to one, impose a more balanced assignment structure and have been proposed as a way to stabilize attention and reduce token collapse~\cite{sander2022sinkformers}. In data-limited biomedical settings, such constraints may regularize token interactions by encouraging more balanced use of image-derived information. Whether this bias improves molecular prediction from pathology images, and for which genes or cancer types, remains an empirical question.

Quantum computing provides a distinct route for generating structured probability distributions and matrix transformations. Although current quantum processors are not yet suitable for large-scale, fault-tolerant machine learning, hybrid quantum-classical workflows can evaluate whether quantum subroutines provide useful structure within otherwise classical models~\cite{abughanem2025ibm,chamberland2022building,lanes2025framework}. Such approaches have been explored across optimization, high-energy physics, protein modeling, healthcare and life sciences~\cite{abbas2024challenges,di2024quantum,doga2024perspective,doga2024can,bose2026advancing,flother2025quantum,burch2025towards}. In parallel, transformer architectures have motivated quantum-inspired and hybrid quantum-classical attention models~\cite{guo2024quantum,khatri2024quixer}.

\begin{figure}[!htb]
    \centering
    \includegraphics[width=1.1\linewidth]{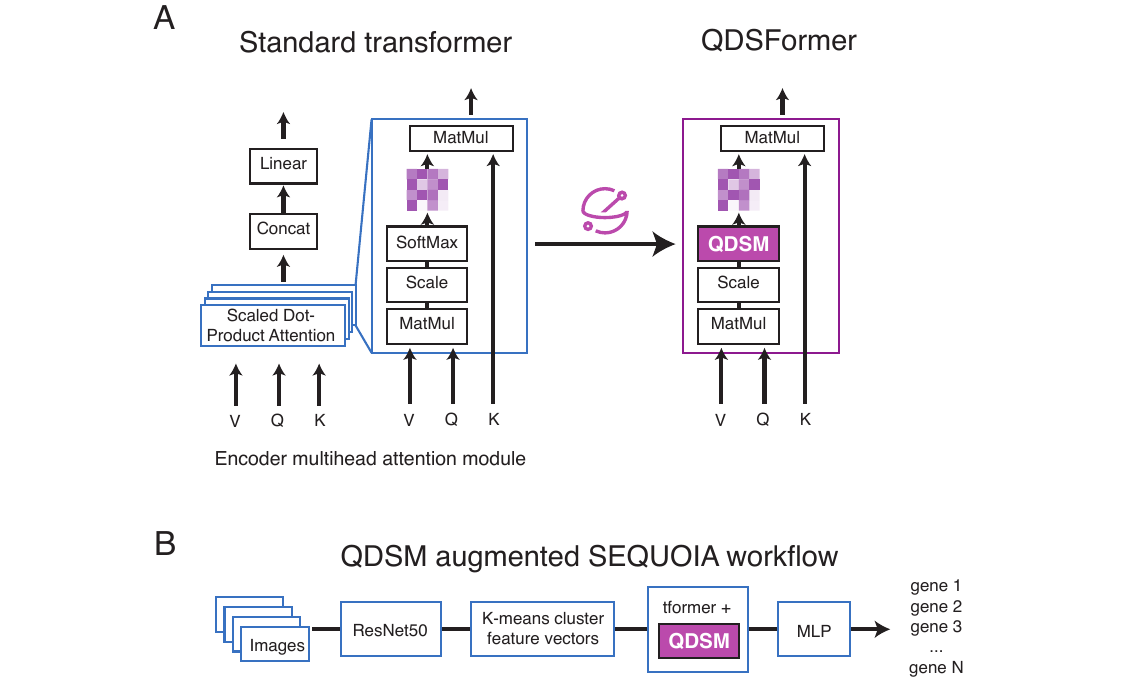}
    \caption{
    \textbf{Quantum-derived doubly stochastic attention integrated into the SEQUOIA histopathology-to-transcriptome workflow.}
    \textbf{a}, Schematic comparison of standard transformer attention and QDSM attention. In a conventional transformer attention head, query and key matrices are used to compute scaled attention scores, which are normalized by a row-wise softmax operation before multiplication with the value matrix. This produces a right-stochastic attention matrix in which each row sums to one. In the QDSM attention head (right), the softmax normalization is replaced by a quantum-derived doubly stochastic matrix generated using the QDSFormer procedure. The resulting attention matrix constrains both rows and columns to sum to one, providing a structured alternative to standard softmax attention while leaving the surrounding transformer architecture unchanged.
    \textbf{b}, SEQUOIA-based workflow for molecular profiling from whole-slide histopathology images. Whole-slide images are tiled, tile-level image features are extracted, and feature vectors are aggregated into representative tokens that are passed to a transformer-based prediction model. The QDSM-augmented model replaces the standard softmax normalization within the transformer attention module with QDSM attention, indicated by the highlighted marker, while preserving the remaining SEQUOIA workflow for transcriptome-wide gene expression prediction.
    }
    \label{fig:overview}
    \vspace{-2mm}
\end{figure}

QDSFormer is a hybrid quantum-classical transformer that replaces softmax attention normalization with a quantum-derived DSM generated using a quantum conditional optimal transport procedure~\cite{born2025quantum,mariella2024quantum}. The quantum circuit produces a DSM that can be inserted into an attention module while leaving the broader transformer architecture classical (Fig.~\ref{fig:overview}a). Prior work showed that QDSFormer can improve performance over classical attention variants in selected vision benchmarks~\cite{born2025quantum}. Its utility in biomedical prediction tasks, where sample sizes are often limited and targets are high-dimensional, has not been systematically evaluated.

Here, we investigate whether quantum-derived doubly stochastic attention can improve histopathology-based molecular profiling. We integrate QDSFormer attention into SEQUOIA by replacing softmax attention with a quantum-derived doubly stochastic matrix (QDSM), while keeping the remaining architecture and prediction task unchanged (Fig.~\ref{fig:overview}b). We evaluate the QDSM-augmented model across 29 cancer types from The Cancer Genome Atlas (TCGA) with matched WSIs and RNA-seq data~\cite{weinstein2013cancer}. We compare full-QDSM training with a lower-overhead warm-start strategy in which QDSM attention is applied only during early training before reverting to classical softmax attention. We further assess cross-cohort behavior in pancreatic cancer using an independent Clinical Proteomic Tumor Analysis Consortium (CPTAC) dataset~\cite{national_cancer_institute_clinical_proteomic_tumor_analysis_consortium_cptac_clinical_2018} and evaluate whether DSMs of relevant sizes can be recovered on IBM quantum processors.

This study addresses whether a quantum-derived doubly stochastic attention bias can yield systematic improvements in molecular prediction from histopathology under specific data conditions. QDSM attention produced gains for subsets of genes and cancer types, with the largest relative improvements in smaller cohorts, while other targets showed reduced performance. These results indicate that QDSM attention changes how image-derived information is used by the model rather than uniformly improving prediction accuracy. Consequently, QDSM attention is best understood as a context- and target-dependent modeling component rather than a drop-in replacement for softmax attention. In settings focused on predefined genes, pathways or biological programs, such selective improvements may support more targeted and data-efficient molecular prediction from histopathology.

\section{Methods}\label{sec11}

\subsection{Patient cohorts and data sources}

We used de-identified whole-slide images (WSIs) and matched bulk RNA-seq profiles from The Cancer Genome Atlas (TCGA), accessed through the Genomic Data Commons Data Portal (\url{https://portal.gdc.cancer.gov/})~\cite{liu2018integrated}. We included cancer types with at least 50 samples containing both diagnostic WSIs and matched transcriptomic measurements. The included cancer types were Adrenocortical carcinoma (ACC), Bladder Urothelial Carcinoma (BLCA), Breast invasive carcinoma (BRCA), Cervical squamous cell carcinoma and endocervical adenocarcinoma (CESC), Colon adenocarcinoma (COAD), Esophageal carcinoma (ESCA), Glioblastoma multiforme (GBM), Head and Neck squamous cell carcinoma (HNSC), Kidney Chromophobe (KICH), Kidney renal clear cell carcinoma (KIRC), Low grade glioma (LGG), Liver hepatocellular carcinoma (LIHC), Lung adenocarcinoma (LUAD), Lung squamous cell carcinoma (LUSC), Mesothelioma (MESO), Ovarian serous cystadenocarcinoma (OV), Pancreatic adenocarcinoma (PAAD), Pheochromocytoma and Paraganglioma (PCPG), Prostate adenocarcinoma (PRAD), Rectum adenocarcinoma (READ), Sarcoma (SARC), Skin Cutaneous Melanoma (SKCM), Stomach adenocarcinoma (STAD), Testicular Germ Cell Tumors (TGCT), Thyroid carcinoma (THCA), Thymoma (THYM), Uterine Corpus Endometrial Carcinoma (UCEC), Uterine Carcinosarcoma (UCS), and Uveal Melanoma (UVM).

Sample counts for each cancer type are shown in Fig.~\ref{fig:tcga}. For each cohort, model training and evaluation were performed using matched WSI and RNA-seq pairs. Gene expression values were modeled as continuous targets, following the SEQUOIA framework~\cite{pizurica2024digital}. Analyses were performed independently for each cancer type to avoid pooling biologically distinct tumor contexts.

For external evaluation, we used an independent pancreatic cancer cohort from the Clinical Proteomic Tumor Analysis Consortium (CPTAC), accessed through the Genomic Data Commons Data Portal. We downloaded 168 pancreatic cancer WSIs with matched gene expression data~\cite{national_cancer_institute_clinical_proteomic_tumor_analysis_consortium_cptac_clinical_2018}. The CPTAC analysis was restricted to the 19,938 genes shared with the TCGA-PAAD dataset. This cohort was used to assess cross-cohort behavior of the QDSM-augmented model after training on TCGA-PAAD.

\subsection{Baseline SEQUOIA model}

SEQUOIA is a transformer-based framework for predicting gene expression profiles from histopathology WSIs~\cite{pizurica2024digital}. In brief, WSIs are tiled, tile-level visual features are extracted using an image encoder, and the resulting feature vectors are aggregated into a fixed number of representative tokens. These tokens are then passed to a vision transformer and multilayer perceptron to predict gene expression values.

In this study, we used the publicly available SEQUOIA implementation, downloaded on January 22, 2025. Unless otherwise stated, we used the default SEQUOIA hyperparameters: 200 training epochs, 10 train-test splits, batch size 16, and random seed 42. Tile-level image features were extracted using a ResNet-50 model pretrained on ImageNet, consistent with the baseline configuration used for comparison. Although the original SEQUOIA study reported stronger performance using the UNI histopathology foundation model, we used ResNet-50 because the licensing terms for UNI restricted its use in the present study. Feature vectors were clustered using $k$-means to generate $K$ aggregated token representations per sample. These aggregated representations served as input tokens to the transformer encoder.

The baseline model used standard scaled dot-product attention with row-wise softmax normalization. The QDSM-augmented model was constructed by replacing the softmax normalization in the attention module with a quantum-derived doubly stochastic matrix, while leaving the remaining SEQUOIA architecture, training configuration, data splits, optimizer settings, and prediction task unchanged. This design allowed the comparison to isolate the effect of the attention normalization.

For TCGA experiments, the number of aggregated clusters $K$ was selected such that the resulting feature dimension was compatible with the QDSM attention implementation. Specifically, $K$ was chosen to yield feature dimensions that were multiples of 8, resulting in values of 48, 64, or 80 depending on the cohort. This enabled construction of an $8 \times 8$ QDSM attention matrix that could be evaluated by classical simulation during model training.

\subsection{Softmax attention}

For an input sequence of $n$ tokens, let
$\mathbf{Q}, \mathbf{K}, \mathbf{V} \in \mathbb{R}^{n \times d}$
denote the query, key, and value matrices. Standard scaled dot-product attention computes the score matrix
\begin{equation}
\mathbf{S} =
\frac{\mathbf{Q}\mathbf{K}^{\top}}{\sqrt{d}},
\end{equation}
and applies a row-wise softmax normalization:
\begin{equation}
\mathbf{A}_{\mathrm{softmax}} =
\mathrm{Softmax}(\mathbf{S}).
\end{equation}
The attention output is then
\begin{equation}
\mathrm{Attn}_{\mathrm{softmax}}(\mathbf{Q},\mathbf{K},\mathbf{V})
=
\mathbf{A}_{\mathrm{softmax}}\mathbf{V}.
\end{equation}

Because the softmax is applied row-wise, each row of $\mathbf{A}_{\mathrm{softmax}}$ sums to one. The resulting matrix is therefore right-stochastic. This normalization constrains the distribution of outgoing attention from each token but does not constrain the total amount of incoming attention assigned to each token across rows.

\subsection{QDSM attention}

We evaluated a hybrid quantum-classical attention mechanism based on QDSFormer~\cite{born2025quantum}. QDSFormer replaces row-wise softmax normalization with a quantum-derived doubly stochastic matrix (QDSM). A doubly stochastic matrix has non-negative entries and satisfies both row- and column-sum constraints:
\begin{equation}
\sum_j A_{ij} = 1
\quad \mathrm{and} \quad
\sum_i A_{ij} = 1.
\end{equation}
This constraint encourages balanced assignment between tokens and provides an alternative attention normalization to softmax.

Given the unnormalized attention scores
\begin{equation}
\mathbf{S} =
\frac{\mathbf{Q}\mathbf{K}^{\top}}{\sqrt{d}},
\end{equation}
QDSFormer maps $\mathbf{S}$ to a doubly stochastic attention matrix:
\begin{equation}
\mathbf{A}_{\mathrm{QDSM}} =
\mathcal{Q}(\mathbf{S}),
\end{equation}
where $\mathcal{Q}(\cdot)$ denotes the quantum conditional optimal transport procedure used to generate the QDSM~\cite{mariella2024quantum,born2025quantum}. The QDSM attention output is then
\begin{equation}
\mathrm{Attn}_{\mathrm{QDSM}}(\mathbf{Q},\mathbf{K},\mathbf{V})
=
\mathbf{A}_{\mathrm{QDSM}}\mathbf{V}.
\end{equation}

In all SEQUOIA training experiments, QDSM attention was evaluated by classical simulation of the corresponding quantum circuit. The transformer model, regression head, loss computation, and optimization were executed classically. Thus, the training experiments evaluate the modeling effect of the QDSM attention normalization, not end-to-end training on a quantum processor.

For the main TCGA experiments, the QDSM attention matrix was restricted to size $8 \times 8$ for computational tractability. The corresponding QDSM was implemented using a parameterized quantum circuit with six data qubits and eight ancilla qubits, resulting in a 14-qubit circuit. Unless otherwise stated, the circuit used eight layers. This configuration was selected to enable repeated training across cancer cohorts and train-test splits while preserving the doubly stochastic structure of the QDSM attention module.

\subsection{Survival analysis}

Overall survival analysis was performed in the ACC cohort using the subset of genes that showed improved image-based prediction under QDSM attention. Survival time was defined as \textit{days\_to\_death} for patients with an event and \textit{days\_to\_last\_followup} for censored patients, with \textit{vital\_status} used to derive the event indicator (1 = death, 0 = censored). Expression values were transformed as \(\log_2(x + 1)\). For each improved gene, a univariable Cox proportional hazards model was fitted using the transformed expression value as a continuous predictor, and hazard ratios (HRs), 95\% confidence intervals, Wald p-values, and Benjamini--Hochberg FDR-adjusted p-values were calculated across all tested genes. Genes were ranked by Cox p-value and summarized in a forest plot. Kaplan--Meier curves were generated for selected genes by dichotomizing patients into high- and low-expression groups using the cohort median transformed expression value, and group differences were assessed with a two-sided log-rank test. This analysis was intended to determine whether genes preferentially improved by QDSM attention also captured clinically relevant prognostic information in ACC.

\subsection{Full-QDSM and warm-start training strategies}

We evaluated two strategies for incorporating QDSM attention into SEQUOIA. In the full-QDSM setting, the softmax normalization in the transformer attention module was replaced by QDSM attention for all 200 training epochs. In the warm-start setting, QDSM attention was used only during the initial phase of training and was then replaced by standard softmax attention for the remaining epochs.

The primary warm-start configuration used QDSM attention for the first five epochs. This strategy was motivated by the hypothesis that early QDSM attention may influence the optimization trajectory while reducing the computational overhead associated with QDSM simulation or future QPU execution. To further characterize this trade-off, we performed an additional warm-start analysis in ACC using QDSM attention for $n \in \{1,3,5,10,50,100\}$ initial epochs before reverting to softmax attention for the remainder of training.

\subsection{Gene-level modeling of QDSM benefit}
To characterize which genes were more likely to benefit from QDSM attention, we modeled the gene-level change in predictive performance as a function of baseline predictivity and molecular features. For each cancer type \(c\), gene \(g\), and train--test split \(s\), we defined the split-level effect size as
\begin{equation}
\Delta r_{g,s,c} = r^{\mathrm{QDSM}}_{g,s,c} - r^{\mathrm{softmax}}_{g,s,c},
\end{equation}
where \(r^{\mathrm{QDSM}}_{g,s,c}\) and \(r^{\mathrm{softmax}}_{g,s,c}\) denote the Pearson correlations between predicted and measured expression under QDSM and softmax attention, respectively. Split-level effect sizes were aggregated across the \(S=10\) train--test splits to obtain a per-gene, per-cancer summary statistic,
\begin{equation}
\Delta r_{g,c} = \frac{1}{S}\sum_{s=1}^{S}\Delta r_{g,s,c}.
\end{equation}
In parallel, baseline predictivity for each gene was defined as the mean Pearson correlation under the softmax-attention model,
\begin{equation}
r^{\mathrm{softmax}}_{g,c} = \frac{1}{S}\sum_{s=1}^{S} r^{\mathrm{softmax}}_{g,s,c}.
\end{equation}

For each gene within each cancer type, we computed molecular covariates intended to capture expression abundance, variability and network context. Mean expression and expression variance were calculated across samples within the corresponding cancer cohort and log-transformed using \(\log(1+x)\):
\begin{equation}
x^{(1)}_{g,c} = r^{\mathrm{softmax}}_{g,c},
\end{equation}
\begin{equation}
x^{(2)}_{g,c} = \log\!\bigl(1 + \overline{E}_{g,c}\bigr),
\end{equation}
\begin{equation}
x^{(3)}_{g,c} = \log\!\bigl(1 + \mathrm{Var}(E_{g,c})\bigr),
\end{equation}
where \(\overline{E}_{g,c}\) and \(\mathrm{Var}(E_{g,c})\) denote the mean and variance of expression for gene \(g\) across samples in cancer type \(c\). To capture transcriptional network context, we computed a within-cancer co-expression degree,
\begin{equation}
x^{(4)}_{g,c} = \mathrm{deg}_{g,c},
\end{equation}
defined as the number of genes whose absolute Pearson correlation with gene \(g\) exceeded a predefined threshold in the within-cancer gene--gene co-expression matrix. In the current implementation, this threshold was set to 0.5.

Before regression, predictor variables were standardized \emph{within cancer type}, such that each predictor was centered and scaled relative to the distribution of that predictor within the corresponding cohort:
\begin{equation}
\tilde{x}^i_{g,c} = \frac{x^{(i)}_{g,c} - \mu_{i,c}}{\sigma_{i,c}},
\end{equation}
where \(\mu_{i,c}\) and \(\sigma_{i,c}\) are the within-cancer mean and standard deviation of predictor \(i\). This transformation preserves relative gene-level position within each cancer type while reducing scale differences among predictors. 

These predictors were used in a mixed-effects regression model:
\begin{equation}
\Delta r_{g,c}
=
\beta_0
+
\beta_1 \tilde{x}^1_{g,c}
+
\beta_2 \tilde{x}^2_{g,c}
+
\beta_3 \tilde{x}^3_{g,c}
+
\beta_4 \tilde{x}^4_{g,c}
+
\epsilon_{g,c},
\end{equation}
where \(\beta_0\) is the fixed intercept, \(\beta_1,\ldots,\beta_4\) are fixed-effect coefficients, and \(\epsilon_{g,c}\) is the residual error term. This model was used to test whether QDSM-associated improvement was related to baseline predictivity, expression abundance, expression variability or co-expression connectivity.

To avoid optimistic in-sample interpretation of residuals, residual-based analyses were performed using \emph{cross-fitted} predictions obtained by leave-one-cancer-out (LOCO) cross-validation. In each LOCO fold, one cancer type was held out, predictor standardization parameters were estimated from the remaining cancers, and the mixed-effects model was fit on the training cancers only. The fitted model was then used to generate predictions for the held-out cancer. Because the cancer-specific random intercept is not identifiable for an unseen cancer, LOCO predictions for the held-out cancer were based on the fixed-effects component alone:
\begin{equation}
\widehat{\Delta r}^{\mathrm{LOCO}}_{g,c}
=
\hat{\beta}_0
+
\hat{\beta}_1 x^{(1),z}_{g,c}
+
\hat{\beta}_2 x^{(2),z}_{g,c}
+
\hat{\beta}_3 x^{(3),z}_{g,c}
+
\hat{\beta}_4 x^{(4),z}_{g,c}.
\end{equation}
Cross-fitted residuals were then computed as
\begin{equation}
\hat{\epsilon}^{\mathrm{LOCO}}_{g,c}
=
\Delta r_{g,c} - \widehat{\Delta r}^{\mathrm{LOCO}}_{g,c}.
\end{equation}
Genes with positive residuals were interpreted as improving more than expected under QDSM attention given their baseline molecular features, whereas genes with negative residuals improved less than expected or worsened relative to expectation.

Residual-based filtering was used to define gene sets for downstream biological interpretation. Because pooled thresholding across all gene--cancer pairs yielded large and poorly interpretable residual-based gene sets, residual filtering was performed separately within each cancer type. Genes were classified as \emph{more improved than expected} if their cross-fitted residual exceeded the cancer-specific upper residual percentile threshold and their observed \(\Delta r_{g,c}\) was positive. Genes were classified as \emph{less improved than expected} if their cross-fitted residual fell below the cancer-specific lower residual percentile threshold and their observed \(\Delta r_{g,c}\) was negative. These cancer-specific residual gene sets were then analyzed by pathway enrichment.

To rank genes by their expected sensitivity to QDSM attention across cancers, we defined a continuous \emph{QDSM benefit score} as the fixed-effects-only prediction from the LOCO procedure,
\begin{equation}
\mathrm{QDSM\ benefit\ score}^{\mathrm{LOCO}}_{g,c}
=
\widehat{\Delta r}^{\mathrm{LOCO}}_{g,c}.
\end{equation}
By construction, this score reflects the expected QDSM benefit attributable to gene-level predictors alone and can therefore be evaluated on held-out cancers. Held-out validation of the QDSM benefit score was performed by assessing the association between \(\mathrm{QDSM\ benefit\ score}^{\mathrm{LOCO}}_{g,c}\) and observed \(\Delta r_{g,c}\) across LOCO folds, including decile-based summaries of observed \(\Delta r\) as a function of the held-out benefit score.

Pathway enrichment analysis (using \textit{gprofiler-official} v1.0.0)  was performed separately on genes classified as more improved than expected and less improved than expected. To improve interpretability and avoid pathways driven by very small or very large gene sets, enrichment analyses were restricted to gene sets containing more than 10 genes and fewer than 300 genes. These analyses were intended to identify biological processes whose QDSM-associated behavior could not be fully explained by baseline predictivity and expression-derived covariates alone.

\subsection{Quantum hardware experiments}

To assess the feasibility of implementing the QDSM-generating primitive on quantum hardware, we performed separate hardware experiments focused on recovery of doubly stochastic matrices. These experiments were independent of SEQUOIA model training. They were designed to test whether the quantum conditional optimal transport circuit used by QDSM attention could recover DSMs on IBM quantum processors.

We evaluated DSMs of size $8 \times 8$, $64 \times 64$, and $128 \times 128$. Circuits used a centrosymmetric ansatz~\cite{mariella2024quantum} with varying expressivity. Specifically, we tested circuits with 8, 4, and 2 layers for the $8 \times 8$, $64 \times 64$, and $128 \times 128$ DSMs, respectively. Experiments were run on IBM Eagle R3 (\textit{ibm\_brisbane}) and IBM Heron R1 (\textit{ibm\_torino}) QPUs. For each configuration, we evaluated 5,000, 10,000, 50,000, and 100,000 shots.

All quantum circuits were transpiled for the target backend using optimization level 3. During hardware execution, error mitigation and error suppression were enabled through dynamical decoupling and Pauli twirling. Dynamical decoupling used an \(X_pX_m\) sequence, and Pauli twirling was applied to reduce coherent error contributions by randomizing them into an effectively stochastic error channel. These mitigation and suppression settings were applied consistently across matrix sizes, shot budgets, and hardware backends.

Hardware-derived DSMs were compared with exact classical calculations and with noise-free statevector simulations. The exact calculation served as the reference solution. The statevector simulation provided an idealized reference for the corresponding circuit in the absence of sampling noise and hardware error. Hardware results reflect finite-shot sampling, residual device noise, readout error, gate error, compilation constraints, and the effects of the applied error suppression and mitigation procedures.

\subsection{Molecular profiling performance}

For each cancer type, model performance was evaluated at the gene level using Pearson correlation between predicted and measured expression values across held-out samples. Pearson correlation was computed separately for each gene and each train-test split. This yielded a distribution of 10 Pearson correlation values per gene for each model configuration.

To compare QDSM attention with standard softmax attention, we performed a two-sided $t$-test for each gene using the 10 split-level Pearson correlations from the QDSM model and the corresponding 10 split-level correlations from the softmax baseline. Genes with nominal $p < 0.05$ and a positive test statistic were classified as improved under QDSM attention. Genes with nominal $p < 0.05$ and a negative test statistic were classified as worsened under QDSM attention.

These gene-level tests were used to compare relative model behavior and to define gene sets for downstream enrichment analysis. Because the tests were performed across many genes, nominal gene-level significance was interpreted as a screening criterion rather than as definitive evidence of individual gene-level improvement. Results were summarized primarily at the level of counts, ratios, and enriched gene sets.

For each cancer type, we computed the number of improved and worsened genes and summarized the net effect using the log-ratio:
\begin{equation}
\log_2
\left(
\frac{N_{\mathrm{improved}}}{N_{\mathrm{worsened}}}
\right),
\end{equation}
where $N_{\mathrm{improved}}$ and $N_{\mathrm{worsened}}$ are the numbers of genes classified as improved or worsened, respectively. If either count was zero, a small pseudocount was added before computing the ratio.

\subsection{Gene set enrichment analysis}

Gene set enrichment analysis was performed separately for genes classified as improved and worsened under QDSM attention. Enrichment was performed using \textit{gprofiler-official} v1.0.0 and STRINGdb~\cite{szklarczyk2023string}. The queried gene set collections included Gene Ontology Biological Process~\cite{ashburner2000gene}, KEGG~\cite{kanehisa2000kegg}, and Reactome~\cite{milacic2024reactome}.

For each cancer type, improved and worsened gene lists were analyzed independently. Enrichment $p$ values were corrected for multiple testing using the Bonferroni method, and corrected $p < 0.01$ was considered significant. Enrichment results were used to determine whether genes whose prediction improved or worsened under QDSM attention were associated with coherent biological processes or pathways. These analyses were interpreted as hypothesis-generating and were not used to infer causal mechanisms or clinical actionability.

\subsection{Association between cohort size and QDSM performance}

To evaluate whether QDSM performance varied with cohort size, we compared per-cancer sample counts with the log-ratio of improved to worsened genes. We also compared sample counts with the log-ratio of enriched gene sets among improved versus worsened genes. Associations were assessed using Pearson correlation across cancer types. These analyses were intended to summarize cohort-level trends and were interpreted cautiously, as cohort size may be correlated with additional factors including tumor heterogeneity, tissue quality, baseline predictability, feature distribution, and transcriptomic variability.

\subsection{Similarity metrics for quantum-derived DSMs}

To evaluate agreement between quantum-derived DSMs and exact classical DSMs, we used Spearman rank correlation and normalized Frobenius distance.
Spearman rank correlation was used to assess whether the relative ordering of matrix elements was preserved. This metric is relevant for attention matrices because downstream model behavior may depend on the relative weighting of tokens, not only on exact element-wise equality. Spearman correlation was computed between the vectorized recovered DSM and the vectorized exact DSM.

Normalized Frobenius distance was used to quantify element-wise deviation between two matrices. For matrices $A$ and $B$, the normalized Frobenius distance was defined as
\begin{equation}
\frac{\|A-B\|_{F}}{N}
=
\frac{\sqrt{\sum_{i,j}(A_{ij}-B_{ij})^2}}{N},
\end{equation}
where $N$ is the number of matrix elements. Normalization by $N$ was used to facilitate comparison across DSMs of different sizes.

\subsection{Software and reproducibility}

SEQUOIA experiments were performed using the publicly available SEQUOIA implementation downloaded on January 22, 2025~\cite{pizurica2024digital}. QDSM attention was implemented by modifying the SEQUOIA attention module to replace row-wise softmax normalization with the QDSFormer QDSM mapping~\cite{born2025quantum}. Unless otherwise stated, all model hyperparameters, train-test split settings, and random seeds were kept fixed between baseline and QDSM-augmented models.

Quantum circuit simulations and hardware experiments were performed using Qiskit (\textit{v.1.4.2}) and IBM quantum backends described above.

\section{Results}\label{sec2}

\subsection{Quantum-derived doubly stochastic attention enabled histopathology-based molecular profiling}

To evaluate whether hybrid quantum-classical attention could support image-based molecular profiling from routine whole-slide histopathology, we integrated a quantum-derived doubly stochastic attention module into the SEQUOIA transformer framework while preserving the broader digital pathology pipeline~\cite{pizurica2024digital,born2025quantum}. In the baseline SEQUOIA configuration, token-token attention is normalized using a row-wise softmax. In the QDSM-augmented model, this normalization was replaced with QDSM, while keeping the remaining model architecture, training procedure and prediction task unchanged. This design allowed us to evaluate the contribution of the attention normalization itself, rather than changes in image preprocessing, feature extraction or downstream regression.

All models were trained using identical settings, including 200 epochs, batch size 16, and 10 train-test splits. Because repeated end-to-end training with large quantum-derived attention matrices on current quantum devices remains computationally expensive, QDSM attention was classically simulated during model training and restricted to an $8 \times 8$ attention matrix. The corresponding QDSM was generated from a 14-qubit circuit with a two-qubit gate depth of 33. We evaluated two QDSM training strategies. In the first, QDSM attention was applied throughout all 200 training epochs. In the second, QDSM attention was used only during early training as a warm start before reverting to standard softmax attention. The warm-start strategy was designed to test whether early exposure to QDSM attention could influence optimization while reducing the overhead associated with QDSM simulation or, in future implementations, quantum hardware execution.

\begin{figure}[!htbp]
    \centering
    \includegraphics[width=1\linewidth]{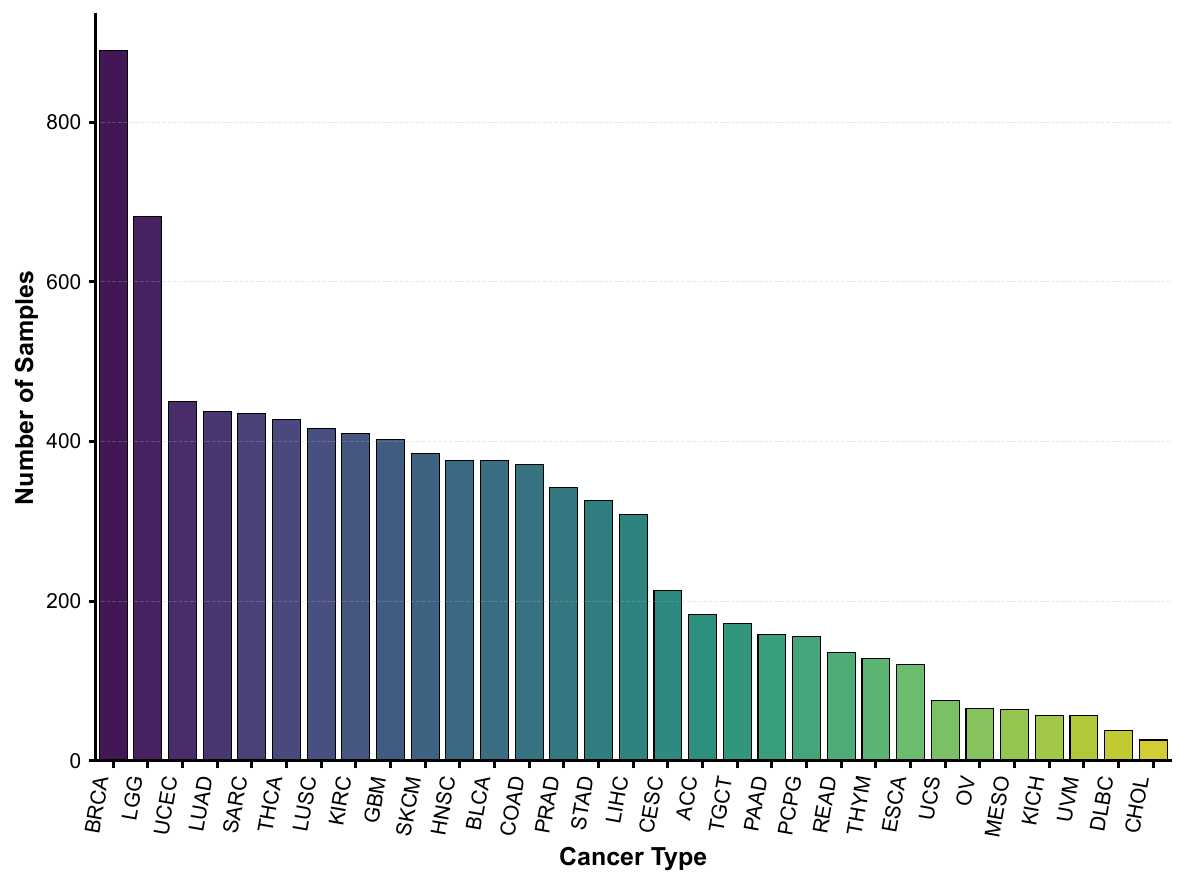}
    \caption{
    \textbf{Number of whole-slide images with matched gene-expression data per cancer type.}
    Shown are the numbers of TCGA samples, defined here as whole-slide histopathology images (WSIs) with matched RNA-seq expression profiles, for cancer types with at least 50 matched WSI--expression pairs. Cancers include: Adrenocortical carcinoma (ACC), Bladder Urothelial Carcinoma (BLCA), Breast invasive carcinoma (BRCA), Cervical squamous cell carcinoma and endocervical adenocarcinoma (CESC), Colon adenocarcinoma (COAD), Esophageal carcinoma (ESCA), Glioblastoma multiforme (GBM), Head and Neck squamous cell carcinoma (HNSC), Kidney Chromophobe (KICH), Kidney renal clear cell carcinoma (KIRC), Low grade glioma (LGG), Liver hepatocellular carcinoma (LIHC), Lung adenocarcinoma (LUAD), Lung squamous cell carcinoma (LUSC), Mesothelioma (MESO), Ovarian serous cystadenocarcinoma (OV), Pancreatic adenocarcinoma (PAAD), Pheochromocytoma and Paraganglioma (PCPG), Prostate adenocarcinoma (PRAD), Rectum adenocarcinoma (READ), Sarcoma (SARC), Skin Cutaneous Melanoma (SKCM), Stomach adenocarcinoma (STAD), Testicular Germ Cell Tumors (TGCT), Thyroid carcinoma (THCA), Thymoma (THYM), Uterine Corpus Endometrial Carcinoma (UCEC), Uterine Carcinosarcoma (UCS), and Uveal Melanoma (UVM).
    }
    \label{fig:tcga}
\end{figure}

\subsection{Hybrid attention selectively improves molecular profiling in data-limited cancer cohorts}

We next assessed how hybrid attention altered image-based molecular profiling across 29 TCGA cancer cohorts by comparing gene-level prediction performance between the QDSM-augmented model and the standard softmax-attention baseline (Fig.~\ref{fig:tcga}). For each cancer type, we trained the baseline SEQUOIA ViT with standard softmax attention and compared it with a QDSM-augmented variant in which QDSM attention replaced softmax normalization throughout training. Model performance was assessed at the gene level using Pearson correlation between predicted and measured expression values across held-out samples. For each gene, the distribution of Pearson correlations across 10 train-test splits was compared between QDSM and softmax attention using a two-sided $t$-test. Genes with nominal $p < 0.05$ and a positive test statistic were classified as improved under QDSM attention, whereas genes with nominal $p < 0.05$ and a negative test statistic were classified as worsened. These gene-level tests were used to compare model behavior across attention mechanisms and should be interpreted as comparative screening statistics rather than validated biomarker findings.

\begin{figure}[!htbp]
\centering
\includegraphics[width=1\textwidth]{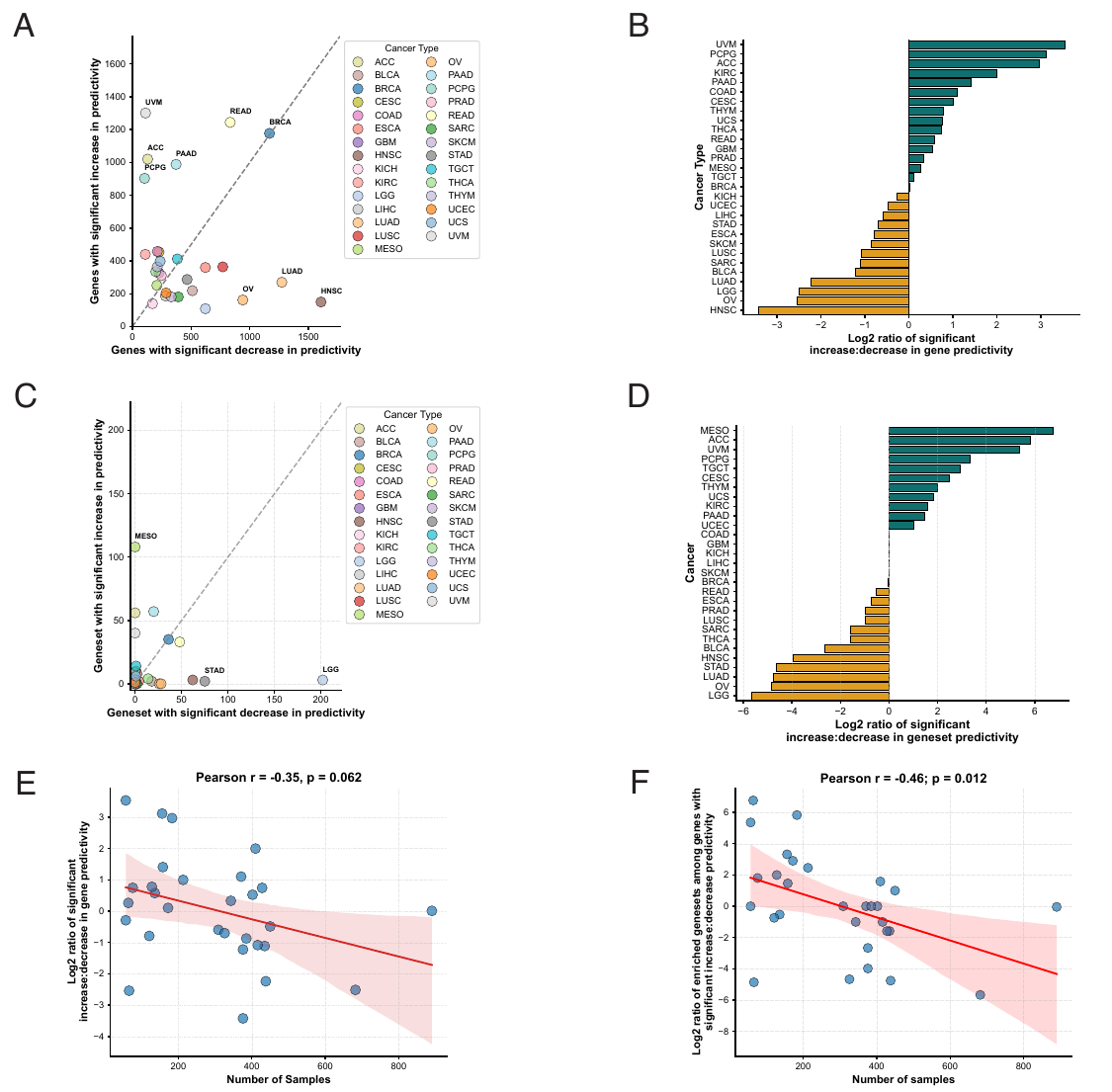}
    \caption{
    \textbf{QDSM attention alters molecular profiling across TCGA cancer types.}
    SEQUOIA ViT models were trained for 200 epochs with either standard softmax attention or QDSM attention in each cancer type with at least 50 samples. Prediction performance was assessed by Pearson correlation between predicted and measured expression across 10 train-test splits.
    \textbf{a}, Number of genes with significantly improved and significantly worsened prediction under QDSM attention relative to softmax attention. Cancer types with the largest net differences are labeled.
    \textbf{b}, $\log_2$ ratio of genes with significantly improved to significantly worsened prediction.
    \textbf{c}, Number of enriched gene sets among genes with significantly improved and significantly worsened prediction. Cancer types with the largest net differences are labeled.
    \textbf{d}, $\log_2$ ratio of enriched gene sets among genes with significantly improved to significantly worsened prediction.
    \textbf{e}, Cohort sample size versus the $\log_2$ ratio of significantly improved to significantly worsened genes, with Pearson correlation and $p$ value reported.
    \textbf{f}, Cohort sample size versus the $\log_2$ ratio of enriched gene sets among significantly improved versus significantly worsened genes, with Pearson correlation and $p$ value reported.
    }
    \label{fig:sig_genes}
\end{figure}

QDSM attention produced heterogeneous effects across cancer types and genes. In 16 of 29 cancer types, more genes showed improved prediction than worsened prediction under QDSM attention relative to softmax attention (Fig.~\ref{fig:sig_genes}a,b). The largest net improvements were observed in UVM, PCPG, ACC, KIRC, and PAAD. However, QDSM attention was not uniformly beneficial: several cancer types showed comparable or greater numbers of worsened genes. These findings indicate that QDSM attention changes the distribution of predictive performance across the transcriptome rather than uniformly improving all gene predictions.

We next examined whether genes with altered predictive performance were enriched for annotated biological pathways. Gene set enrichment was performed separately for genes that improved and those worsened under QDSM attention as compared to softmax attention. Several cancer types showed a larger number of enriched gene sets among improved genes than among worsened genes, including ACC, UVM, PCPG, MESO, CESC, and UCS (Fig.~\ref{fig:sig_genes}c,d). In contrast, other cancers, including THCA, LUAD, BLCA, HNSC, STAD, OV, and LGG, showed enrichment patterns that were more prominent among worsened genes. Thus, the effects of QDSM attention were not only gene-specific but also pathway- and cancer-type-specific. This pattern suggests that QDSM attention may preferentially improve the prediction of certain molecular programs while reducing performance for others.

We also assessed whether the relative benefit of QDSM attention was associated with cohort size. Across cancer types, the ratio of improved to worsened genes showed a negative association with sample size ($r=-0.35$, $p=0.062$; Fig.~\ref{fig:sig_genes}e). A stronger negative association was observed for the ratio of enriched gene sets among improved versus worsened genes ($r=-0.42$, $p=0.012$; Fig.~\ref{fig:sig_genes}f). These results are consistent with the hypothesis that QDSM attention may be more useful in smaller cohorts~\cite{caro2022generalization}, although cohort size is only one of several factors that may influence performance. Tumor biology, tissue heterogeneity, baseline predictability, and transcriptomic variability may also contribute to the observed differences across cancer types.

\subsection{Hybrid attention yields modest transcriptome-wide effects but improves specific molecular targets
}

To complement significance-based comparisons, we quantified the change in gene-level prediction correlation produced by hybrid attention and found that transcriptome-wide shifts were generally modest, with larger gains concentrated in selected molecular targets. We evaluated this change in prediction using a gene-level prediction correlation improvement (PCI) of QDSM attention as
\(\Delta r = r_{\mathrm{QDSM}} - r_{\mathrm{softmax}}\), where \(r\) denotes the Pearson correlation between predicted and measured expression (Fig.~\ref{fig:effect_size}, Table~\ref{tab:deltar_effect_sizes}). Across cancer types, the genome-wide median \(\Delta r\) values were generally modest and centered near zero, indicating that QDSM attention did not uniformly shift image-based molecular profiling accuracy. The largest positive median shifts were observed in UVM (\(\mathrm{median}\ \Delta r = 0.15\)), PCPG (\(\mathrm{median}\ \Delta r = 0.06\)), ACC (\(\mathrm{median}\ \Delta r = 0.06\)) and KIRC (\(\mathrm{median}\ \Delta r = 0.04\)), whereas the largest negative shifts were observed in DLBC, OV, KICH and HNSC.

Although the genome-wide median effects were modest, genes classified as significantly improved under QDSM attention showed larger PCIs. For example, the median \(\Delta r\) among improved genes was 0.28 in ACC, 0.25 in PCPG and 0.45 in UVM. Conversely, genes classified as worsened also showed non-trivial negative PCIs in several cohorts. This indicates that QDSM attention does not act as a uniform enhancer of image-based molecular profiling, but instead redistributes predictive performance across genes. The PCI distributions therefore support a target-specific interpretation of QDSM attention, in which benefit depends on the cancer type and molecular targets considered.

\begin{figure}[!htbp]
    \centering
    \includegraphics[width=1\linewidth]{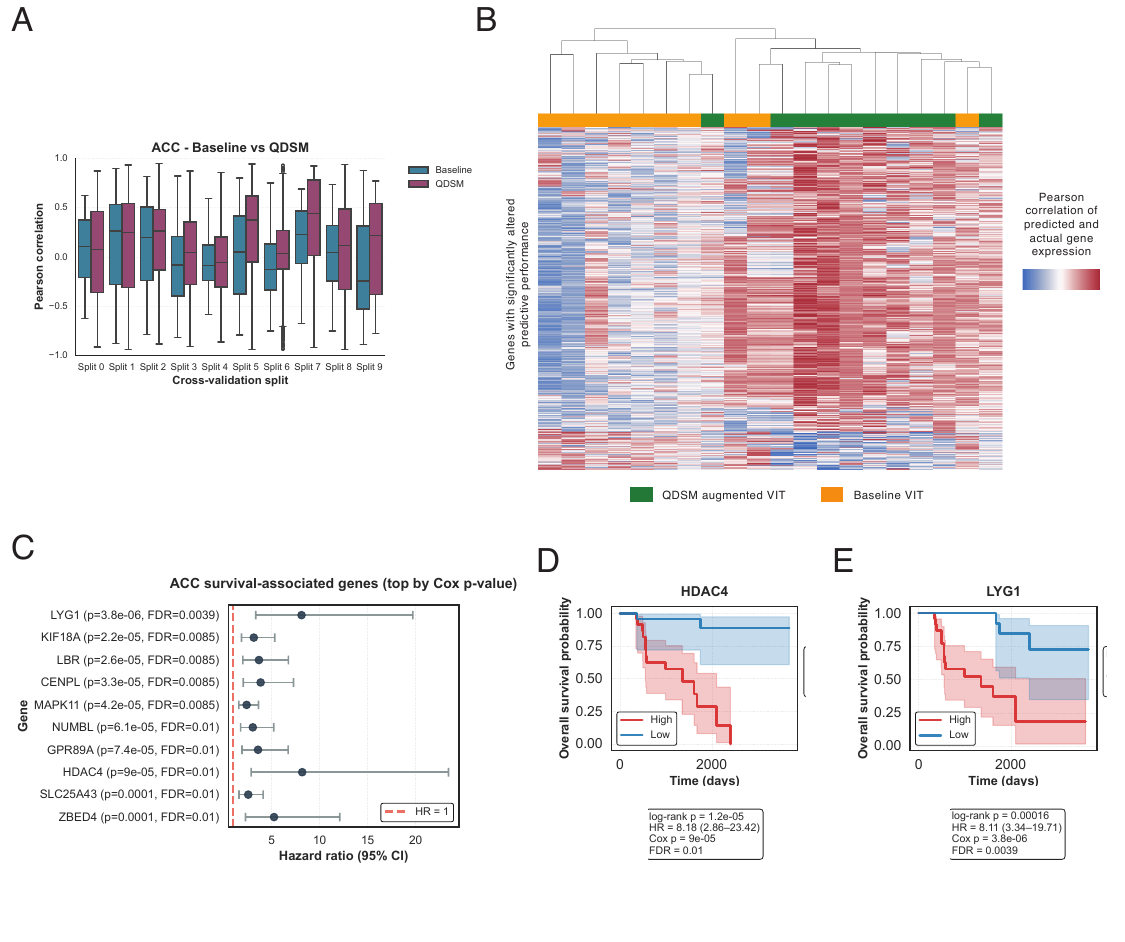}
    \caption{
    \textbf{QDSM attention improves prediction for a subset of genes in ACC.}
    \textbf{a}, Boxplots showing the distribution of gene-level Pearson correlations between predicted and measured expression across 10 train-test splits for baseline SEQUOIA with softmax attention and SEQUOIA with QDSM attention. Higher Pearson correlation values indicate better gene expression prediction.
    \textbf{b}, Heatmap of Pearson correlation values across train-test splits for genes with significantly altered predictive performance under QDSM attention relative to softmax attention, based on a two-sided $t$-test with nominal $p < 0.05$.
    \textbf{c}, Forest plot of the top survival-associated genes in ACC among genes with improved prediction under QDSM attention. Univariable Cox proportional hazards models were fitted for overall survival, and the ten strongest associations are shown. Points indicate hazard ratios (HRs) and horizontal lines indicate 95\% confidence intervals. All genes shown had HR $> 1$ and both nominal $p$ values and FDR values below $10^{-23}$, indicating that higher expression was associated with shorter overall survival.
    \textbf{d,e}, Kaplan--Meier curves for representative survival-associated genes, \textit{HDAC4} (d) and \textit{LYG1} (e) in ACC. Patients were stratified into high- and low-expression groups using the cohort median of the \(\log_2(x)\)-transformed expression value for each gene. Higher expression was associated with worse survival for each gene shown. Insets report the HR, Cox-model $p$ value, FDR, and two-sided log-rank $p$ value.
    }
    \label{fig:ACC_example}
\end{figure}

\subsection{Hybrid attention improves image-based molecular profiling in adrenocortical carcinoma}

Given that adrenocortical carcinoma (ACC) showed one of the strongest net benefits from hybrid attention, we examined this rare and molecularly heterogeneous cancer in greater detail as a representative data-limited use case. ACC is an aggressive endocrine malignancy in which prior multi-omic studies have identified recurrent alterations involving TP53, Wnt/$\beta$-catenin, PKA signaling, cell-cycle regulation, chromosomal instability, and epigenetic control~\cite{zheng_comprehensive_2016,lerario_update_2022,kamilaris_adrenocortical_2020}. This biological heterogeneity makes ACC a useful setting in which to assess whether an alternative attention normalization can improve image-based inference of selected transcriptomic programs from histopathology.

\subsubsection{Hybrid attention produces a net increase in gene-level prediction performance in ACC}

Compared with the baseline SEQUOIA model using softmax attention, the QDSM-augmented model showed a rightward shift in the distribution of gene-level Pearson correlations across the 10 train--test splits, indicating a net improvement in transcriptome-wide prediction performance in ACC (Fig.~\ref{fig:ACC_example}a). At nominal $p < 0.05$, 1,020 genes showed significantly improved prediction under QDSM attention, whereas 130 genes showed worsened prediction (Fig.~\ref{fig:ACC_example}b). This marked imbalance indicates that, in ACC, hybrid attention increased the number of genes whose expression could be more accurately inferred from whole-slide imaging features. At the same time, the presence of worsened genes indicates that the doubly stochastic attention constraint redistributes predictive performance across targets and should not be interpreted as uniformly beneficial for all genes.

\subsubsection{Genes improved by hybrid attention map to ACC-relevant biological programs}

To determine whether the genes improved under hybrid attention reflected coherent biological programs, we performed gene-set enrichment analysis separately on genes with significantly improved and significantly worsened prediction. Among genes with improved prediction, enriched processes included DNA damage and repair, transcriptional and epigenetic regulation, and protein ubiquitination. These annotations are consistent with known features of ACC biology. Integrated genomic analyses have identified frequent disruption of TP53-associated pathways, extensive chromosomal instability, and molecular subtypes characterized by distinct transcriptional and epigenetic programs~\cite{zheng_comprehensive_2016,lerario_update_2022}. Additional whole-genome and transcriptome studies of advanced ACC have highlighted alterations affecting DNA repair and homologous recombination deficiency-related biology, supporting the relevance of DNA-damage programs in at least a subset of tumors~\cite{lavoie_whole-genome_2022}. Likewise, transcriptional regulatory analyses have shown that ACC subtypes can be distinguished by regulon activity patterns associated with prognosis and molecular phenotype~\cite{muzzi_comprehensive_2022}. The enrichment of gene-regulatory terms among genes improved by hybrid attention is therefore compatible with the established molecular stratification of ACC.

The enrichment of ubiquitination-related processes is also biologically plausible in ACC. The TCGA pan-genomic characterization identified recurrent deletions of \textit{ZNRF3}, an E3 ubiquitin ligase and negative regulator of Wnt signaling, and reinforced the importance of Wnt/$\beta$-catenin dysregulation in ACC~\cite{zheng_comprehensive_2016,kamilaris_adrenocortical_2020}. Because ubiquitin-mediated regulation intersects with protein turnover, signaling control, and cell-cycle progression, improved prediction of genes in this category may reflect oncogenic states that are morphologically encoded in tissue architecture. These enrichment results do not establish causality or clinical actionability for individual genes, but they support the conclusion that the genes preferentially improved by hybrid attention are biologically organized rather than randomly distributed.

\subsubsection{Genes preferentially captured by hybrid attention carry adverse prognostic signal in ACC}

To determine whether the molecular targets preferentially captured by hybrid attention also carried prognostic information in ACC, we performed gene-level overall-survival analysis using the subset of genes that showed improved image-based prediction under QDSM attention. Univariable Cox modeling of \(\log_2(x)\)-transformed gene expression values identified a highly significant adverse-prognosis gene set, with the ten strongest associations corresponding to \textit{LYG1}, \textit{KIF18A}, \textit{LBR}, \textit{CENPL}, \textit{MAPK11}, \textit{NUMBL}, \textit{GPR89A}, \textit{SLC25A43}, \textit{ZBED4}, and \textit{HDAC4}, all with hazard ratios greater than 1 and $p$ values below $10^{-4}$ (Fig.~\ref{fig:ACC_example}C). Thus, for each of these genes, higher expression was associated with shorter overall survival. Forest-plot visualization showed consistently elevated hazard ratios across the top-ranked genes, and Kaplan--Meier analysis of representative genes demonstrated clear survival separation between high- and low-expression groups (Fig.~\ref{fig:ACC_example}D,E). In particular, \textit{LYG1} and \textit{HDAC4} showed the largest adverse effect sizes in the top-ranked set, although these stronger associations likely reflect broader coordinated risk programs rather than fully independent effects.

Biologically, this adverse-prognosis signature converged on pathways highly compatible with aggressive ACC biology. \textit{KIF18A}, a mitotic kinesin involved in chromosome alignment and cell-cycle progression, is especially notable because prior pan-cancer analysis has already identified its overexpression as adverse for overall survival in ACC~\cite{liu_comprehensive_2023}. \textit{CENPL} and \textit{KIF18A} both align with mitotic and centromere-associated programs, whereas \textit{LBR} supports a complementary interpretation centered on nuclear-envelope integrity and genomic stability, processes that are relevant to cancers with extensive chromosomal instability~\cite{patil_nuclear_2023}. \textit{HDAC4} implicates chromatin and epigenetic regulation, a biologically plausible axis in view of the transcriptional and epigenetic subclasses already described in ACC~\cite{cuttini_hdac4_2023}, and is the target of several therapeutics~\cite{hassell_histone_2019}. \textit{MAPK11}, \textit{NUMBL}, and \textit{GPR89A} point toward broader stress-signaling, developmental, and membrane-trafficking programs that have been implicated in cancer progression outside ACC, whereas \textit{SLC25A43} suggests a mitochondrial or metabolic component and \textit{ZBED4} is consistent with emerging cross-cancer prognostic biology~\cite{katopodis_p38_2021,garcia_heredia_numb_2018,gabrielson_mitochondrial_2013, ding_zbed4_2025}. Finally, \textit{LYG1} is particularly interesting because prior work links it to CD4$^{+}$ T-cell activation and antitumor immune function, raising the possibility that part of the ACC survival signal captured here reflects immune-state variation rather than tumor-cell intrinsic proliferation alone~\cite{liu_lyg1_2021}. Taken together, these findings suggest that the genes preferentially captured by hybrid attention are not only biologically coherent, but also enriched for clinically relevant adverse-prognosis information in ACC.

Taken together, the ACC analysis indicates that QDSM attention can improve prediction of a biologically organized and clinically relevant subset of genes in a rare, molecularly heterogeneous cancer. The genes most strongly associated with QDSM-related benefit were enriched for processes with established relevance to ACC and aggressive cancer states, including cell-cycle progression, genome stability, transcriptional and epigenetic regulation, and stress-response biology, and a subset of these genes also showed strong adverse associations with overall survival. These findings support the view that hybrid attention may be particularly useful when it enhances prediction of molecular programs that are both histopathologically encoded and prognostically informative. In this context, ACC illustrates a target-specific use case in which quantum-derived doubly stochastic attention improves image-based molecular inference for biologically coherent gene sets in a rare cancer while preserving the need for pathway-level validation and broader external confirmation.

\subsection{Early hybrid-attention warm-starting partially preserves molecular profiling gains while reducing training overhead
}
As full hybrid-attention training increases computational cost, we next asked whether the molecular profiling benefit of QDSM attention could be preserved using a lower-overhead training strategy. To test this, we implemented a mixed hybrid-classical schedule in which QDSM attention was applied only during the first five training epochs before reverting to standard softmax attention for the remainder of optimization. This warm-start design was intended to determine whether early exposure to structured attention was sufficient to influence downstream model behavior while reducing the simulation burden associated with applying QDSM attention throughout all 200 training epochs. By comparing this reduced-cost strategy with full-QDSM and softmax-only training across cancer cohorts, we assessed whether the gains from hybrid attention were primarily driven by early optimization dynamics or instead required sustained exposure to the structured attention mechanism.

\begin{figure}[!htbp]
\centering
\includegraphics[width=1\textwidth]{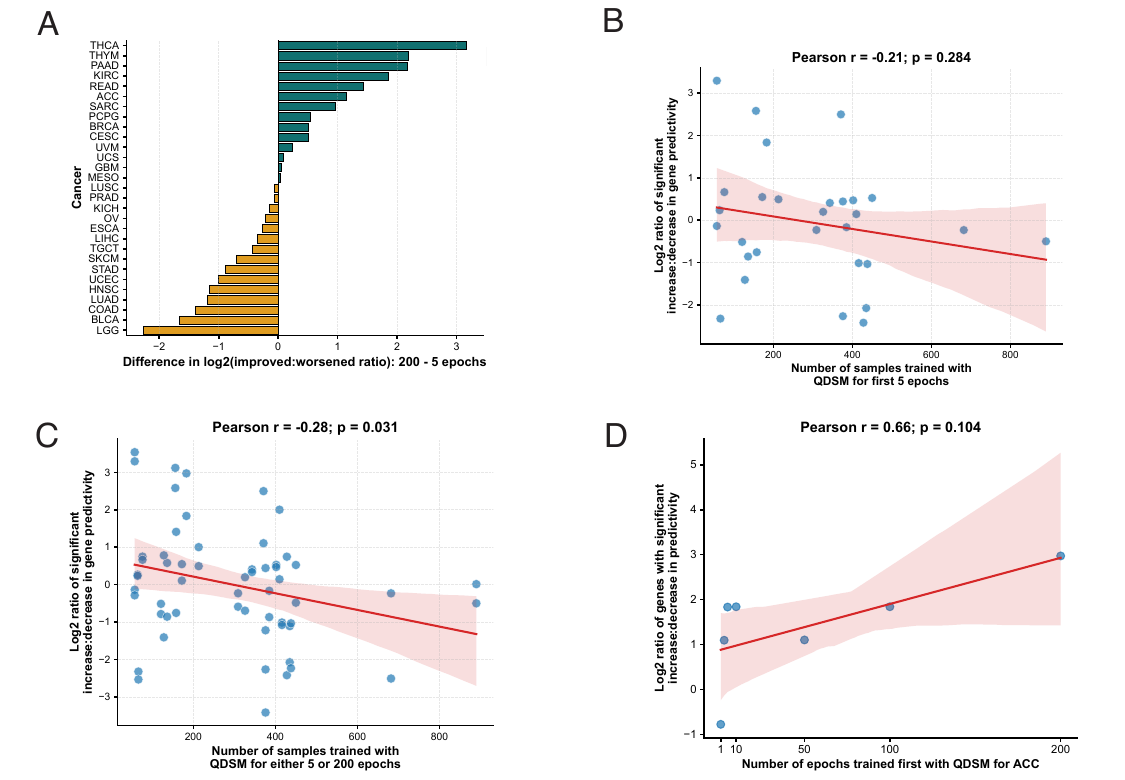}
    \caption{
    \textbf{Performance of mixed QDSM and softmax training.}
    \textbf{a}, Difference in the \(\log_2\) ratio of genes with significantly improved to significantly worsened prediction between full-QDSM training (200 epochs) and QDSM warm-starting (5 epochs).
    \textbf{b}, Cohort sample size versus the \(\log_2\) ratio of significantly improved to significantly worsened genes after 5-epoch QDSM warm-starting, with Pearson correlation and \(p\) value reported.
    \textbf{c}, Cohort sample size versus the \(\log_2\) ratio of significantly improved to significantly worsened genes across both 5-epoch warm-start and 200-epoch full-QDSM training, with Pearson correlation and \(p\) value reported.
    \textbf{d}, QDSM warm-start duration versus the \(\log_2\) ratio of significantly improved to significantly worsened genes in ACC, with Pearson correlation and \(p\) value reported.
    }
    \label{fig:mixed_learning}
\end{figure}

Across the 29 TCGA cancer types, five-epoch QDSM warm-starting produced a net increase in the number of improved versus worsened genes in 14 cancers (Fig.~\ref{fig:mixed_learning}a). The pattern of benefit across cancer types broadly resembled that observed with full-QDSM training, although the magnitude of improvement was generally reduced. Under the five-epoch warm-start configuration alone, the association between cohort size and the ratio of improved to worsened genes remained negative but not statistically significant ($r=-0.21$, $p=0.284$; Fig.~\ref{fig:mixed_learning}b). When warm-start and full-QDSM results were considered together, the association between smaller cohort size and greater relative improvement was statistically significant ($r=-0.28$, $p=0.031$; Fig.~\ref{fig:mixed_learning}c).

To further examine the effect of warm-start duration, we varied the number of initial QDSM epochs in ACC before switching back to softmax attention. We evaluated 1, 3, 5, 10, 50 and 100 QDSM warm-start epochs. Increasing the number of QDSM epochs was associated with larger relative improvement in ACC ($r=0.66$, $p=0.104$), although this analysis was limited to a small number of tested durations and did not reach statistical significance (Fig.~\ref{fig:mixed_learning}d). Warm-start durations of 5--10 epochs achieved substantial, but incomplete, recovery of the full-QDSM effect. These findings suggest that QDSM attention can influence model training early in optimization, but full exposure to QDSM attention may be required to obtain the largest improvements in some cohorts.

\subsection{External pancreatic cancer evaluation shows selective molecular transferability under cross-cohort shift}

To assess whether hybrid attention could support external transfer of image-based molecular profiling, we evaluated the model in an independent pancreatic cancer cohort from the Clinical Proteomic Tumor Analysis Consortium (CPTAC) after initial training in the TCGA-PAAD setting~\cite{national_cancer_institute_clinical_proteomic_tumor_analysis_consortium_cptac_clinical_2018}. This experiment tested whether the structured attention bias introduced by QDSM could remain informative across cohorts that differ in histology preparation, patient composition, and molecular profiling workflows. A model trained on TCGA-PAAD whole-slide images was fine-tuned on CPTAC pancreatic cancer data, and performance was compared between the baseline SEQUOIA model and the QDSM-augmented model across genes shared between the two datasets. This design allowed us to examine whether hybrid attention could preserve benefit for selected molecular targets under external domain shift rather than only within the discovery cohort.

\begin{figure}[!htbp]
\centering
\includegraphics[width=1\textwidth]{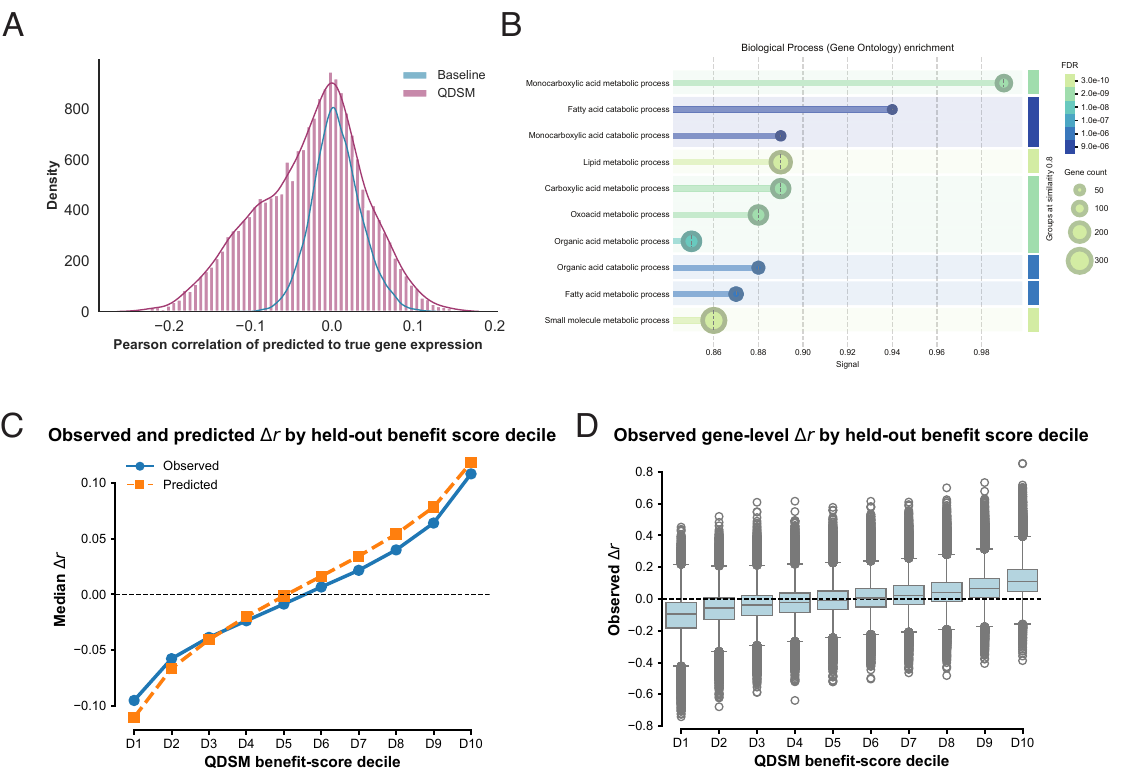}
    \caption{
        \textbf{Cross-cohort transfer performance on CPTAC pancreatic cancer and gene-level modeling of expected QDSM benefit.}
        \textbf{a}, Distribution of gene-level Pearson correlation values between predicted and measured expression in the CPTAC pancreatic cancer cohort for the baseline SEQUOIA model with softmax attention and the QDSM-augmented model.
        \textbf{b}, Pathways enriched according to STRINGdb among genes with significantly improved prediction under QDSM attention in the CPTAC pancreatic cancer cohort.
        \textbf{c}, Observed and predicted median \(\Delta r\) values as a function of QDSM benefit-score decile. Genes were ranked by the leave-one-cancer-out (LOCO)-derived QDSM benefit score and divided into deciles. Both observed and predicted \(\Delta r\) increased across deciles, indicating that genes assigned higher expected QDSM benefit tended to show larger realized gains under QDSM attention.
        \textbf{d}, Distribution of observed gene-level \(\Delta r\) values by QDSM benefit-score decile. Boxplots summarize the spread of observed \(\Delta r\) values within each decile, showing a positive association between the QDSM benefit score and realized improvement under QDSM attention.
    }
    \label{fig:cptac}
\end{figure}

Relative to the softmax-attention baseline, QDSM attention improved prediction for approximately 2,000 genes but worsened prediction for approximately 6,200 genes (Fig.~\ref{fig:cptac}a). These results indicate that the QDSM-induced attention bias can transfer for a subset of genes, but does not provide uniform improvement under cross-cohort domain shift. Differences between TCGA and CPTAC, including cohort composition, tissue processing, slide preparation, sequencing protocols, and tumor heterogeneity, may influence which molecular targets remain predictable across datasets. Thus, QDSM attention should not be interpreted as a general solution to cross-cohort transfer, but rather as a modeling component whose utility depends on the target genes and dataset context.

Among the genes better predicted under QDSM attention, several have established relevance to pancreatic cancer biology. Improved genes included ARID1A, a tumor suppressor frequently altered in pancreatic cancer~\cite{kuo2023arid1a,tomihara2021loss}, and STK11, germline alterations of which are associated with increased pancreatic cancer risk~\cite{su1999germline}. QDSM attention also improved prediction of genes involved in metabolic regulation, including G6PD, ALDOB, CPT1A and SREBF1/2, and metabolic pathways were enriched among improved genes according to STRINGdb analysis (Fig.~\ref{fig:cptac}b)~\cite{szklarczyk2023string}. In addition, lineage-associated transcriptional regulators including GATA6, FOXA2 and SPDEF showed improved prediction. These genes have been implicated in pancreatic ductal adenocarcinoma lineage state or subtype biology~\cite{o2020gata6,milan2019foxa,tonelli2024mucus}. 

These findings suggest that the subset of genes improved under QDSM attention includes biologically interpretable pancreatic cancer programs. However, the larger number of worsened genes underscores the target-specific nature of the effect. For applications focused on predefined pathways or molecular subtypes, such selective improvement may be useful if the genes of interest are among those improved. For transcriptome-wide reconstruction, the observed trade-off indicates that additional model selection, ensembling, or target-specific optimization would be required.

\subsection{Gene-level modeling reveals predictable cancer-specific drivers of hybrid attention benefit}

To determine which molecular targets were most likely to benefit from hybrid attention, we modeled the per-gene change in predictive performance, \(\Delta r = r_{\mathrm{QDSM}} - r_{\mathrm{softmax}}\), as a function of baseline predictivity, expression abundance, expression variance, and co-expression degree across cancer cohorts. In the full mixed-effects model fit to all gene--cancer pairs, the correlation between observed and predicted \(\Delta r\) values was 0.66 (Fig.~\ref{fig:mixed_model}a). When evaluated using leave-one-cancer-out (LOCO) cross-validation, this correlation decreased to 0.52 (Fig.~\ref{fig:mixed_model}b), indicating that part of the stronger in-sample fit reflected cohort-specific structure. Nevertheless, the LOCO analysis showed that baseline gene-level features explained a meaningful fraction of the variation in hybrid-attention benefit, while also indicating that substantial residual, cancer-specific structure remained unexplained.

The LOCO residual analysis further clarified how QDSM benefit varied across cancers. By construction, the global distribution of cross-fitted residuals was centered near zero (Fig.~\ref{fig:mixed_model}c), indicating no systematic directional bias in the held-out predictions. Within individual cancers, residual distributions were broadly symmetric but differed in spread and central tendency. UVM, ACC, PCPG, UCS and KIRC showed the most positive median residual levels, indicating that genes in these cancers tended to improve more than expected under QDSM attention after accounting for baseline gene properties. Thus, these cancers were distinguished by a positive shift in residual \(\Delta r\), rather than consistently by the largest residual variance.

To summarize expected QDSM benefit independent of these residual shifts, we ranked genes by the LOCO-derived \emph{QDSM benefit score} and divided them into deciles. Observed and predicted \(\Delta r\) values increased consistently across these deciles and showed strong concordance (Fig.~\ref{fig:cptac}c), indicating that genes assigned higher expected QDSM benefit also tended to show larger realized gains under QDSM attention. A complementary boxplot analysis showed that the distribution of observed \(\Delta r\) shifted upward across increasing benefit-score deciles (Fig.~\ref{fig:cptac}d), supporting a positive relationship between the modeled benefit score and held-out QDSM-associated improvement. Together, these results indicate that baseline predictivity and expression-derived features capture a meaningful component of QDSM benefit at the gene level, while the residual analysis identifies additional cancer- and pathway-specific effects not fully explained by those predictors.

We then used LOCO residuals to identify genes that improved more or less than expected after accounting for baseline predictivity, expression abundance, expression variance and co-expression degree. To improve interpretability, residual-based pathway analysis was performed using cancer-specific residual thresholds, and pathway sets were restricted to gene sets containing more than 10 genes and fewer than 300 genes. Genes with positive residuals beyond the upper threshold were classified as \emph{more improved than expected}, whereas genes with negative residuals beyond the lower threshold were classified as \emph{less improved than expected}. Pathway enrichment analysis of the more-improved-than-expected genes identified processes related to tissue development, morphogenesis and structural regulation, including negative regulation of cell development, telencephalon development, regulation of neuron apoptotic process, regulation of blood circulation, negative regulation of cell migration, actomyosin structure organization, heart morphogenesis, response to fibroblast growth factor and response to transforming growth factor beta (Fig.~\ref{fig:mixed_model}d). Additional enrichment for regulation of leukocyte mediated immunity suggests that the residual QDSM signal was not restricted to a single biological theme.

By contrast, genes classified as less improved than expected were enriched for pathways related to ribosomal metabolism, DNA replication and repair, and adaptive immune regulation, including rRNA metabolic process, recombinational repair, double-strand break repair via homologous recombination, DNA replication, regulation of adaptive immune response, and positive regulation of leukocyte- and lymphocyte-mediated immunity (Fig.~\ref{fig:mixed_model}d). The presence of immune-related pathways in the less-improved-than-expected set indicates that, once baseline molecular features are taken into account, selected immune programs may benefit less from QDSM attention than other categories of genes. Overall, the residual-based enrichment analysis suggests that QDSM-associated benefit is not fully explained by generic gene properties and that the remaining signal is biologically structured.

Taken together, these analyses indicate that QDSM-associated benefit is partly predictable from baseline gene properties and partly context-specific. The LOCO-derived QDSM benefit score captures a gene-level gradient of expected improvement across cancers, whereas the cross-fitted residuals identify genes and pathways whose observed QDSM-associated behavior is larger or smaller than expected in a given tumor context. The residual-based enrichment results suggest that this structure is biologically meaningful but not uniformly directional: developmental and morphogenetic pathways were enriched among genes that improved more than expected, whereas DNA replication, homologous recombination and adaptive immune pathways were enriched among genes that improved less than expected. The appearance of \emph{regulation of leukocyte mediated immunity} in both sets indicates that QDSM effects are not uniform even within a shared biological category, but likely depend on the specific genes and contexts involved.

\begin{table}
\centering
\caption{\textbf{Quantum circuit configurations used for QPU DSM recovery experiments.}
Circuit sizes evaluated on IBM quantum processors for doubly stochastic matrix (DSM) recovery. For each DSM size, the table reports the number of qubits, circuit layers, two-qubit gate depth, and total number of two-qubit gates after transpilation with optimization level 3.}
\label{tab:qpu_circuit_sizes}
\begin{tabular}{lcccc}
\toprule
\textbf{DSM size} & \textbf{Layers} & \textbf{Qubits} & \textbf{Two-qubit depth} & \textbf{Two-qubit gates} \\
\midrule
$8 \times 8$     & 8 & 14 & 33 & 133 \\
$64 \times 64$   & 4 & 26 & 66 & 310 \\
$128 \times 128$ & 2 & 30 & 40 & 173 \\
\bottomrule
\end{tabular}
\end{table}

\subsection{Quantum hardware reproduces the structured attention primitive underlying hybrid molecular profiling
}

To evaluate the feasibility of the quantum component underlying the hybrid attention mechanism, we performed separate hardware experiments focused on recovering the doubly stochastic matrix primitive on IBM quantum processors. This hardware validation was conducted independently of whole-slide model training because current quantum hardware and deep learning workflows do not yet support efficient end-to-end execution of the full histopathology pipeline on quantum processing units (QPUs). By isolating DSM recovery from model optimization, we aimed to test whether the structured attention primitive required for QDSM attention can be implemented on available hardware, even in the absence of live QPU-based training or inference.

We evaluated DSM construction for matrices of size $8 \times 8$, $64 \times 64$ and $128 \times 128$, corresponding to increasingly large attention normalizations (Table~\ref{tab:qpu_circuit_sizes}). These matrix sizes were chosen to span the small DSM used in the SEQUOIA training experiments and larger DSMs that are more representative of attention modules with greater token capacity. Circuits with varying expressivity were executed on IBM Eagle R3 (\textit{ibm\_brisbane}) and IBM Heron R1 (\textit{ibm\_torino}) devices. Hardware-derived DSMs were compared with exact classical calculations and noise-free statevector simulations. The statevector simulations provide an idealized reference for the circuit in the absence of sampling noise and hardware error, whereas the QPU results reflect finite-shot sampling, gate noise, readout error, and device-specific connectivity constraints.

\begin{figure}[!htbp]
    \centering
    \includegraphics[width=1\linewidth]{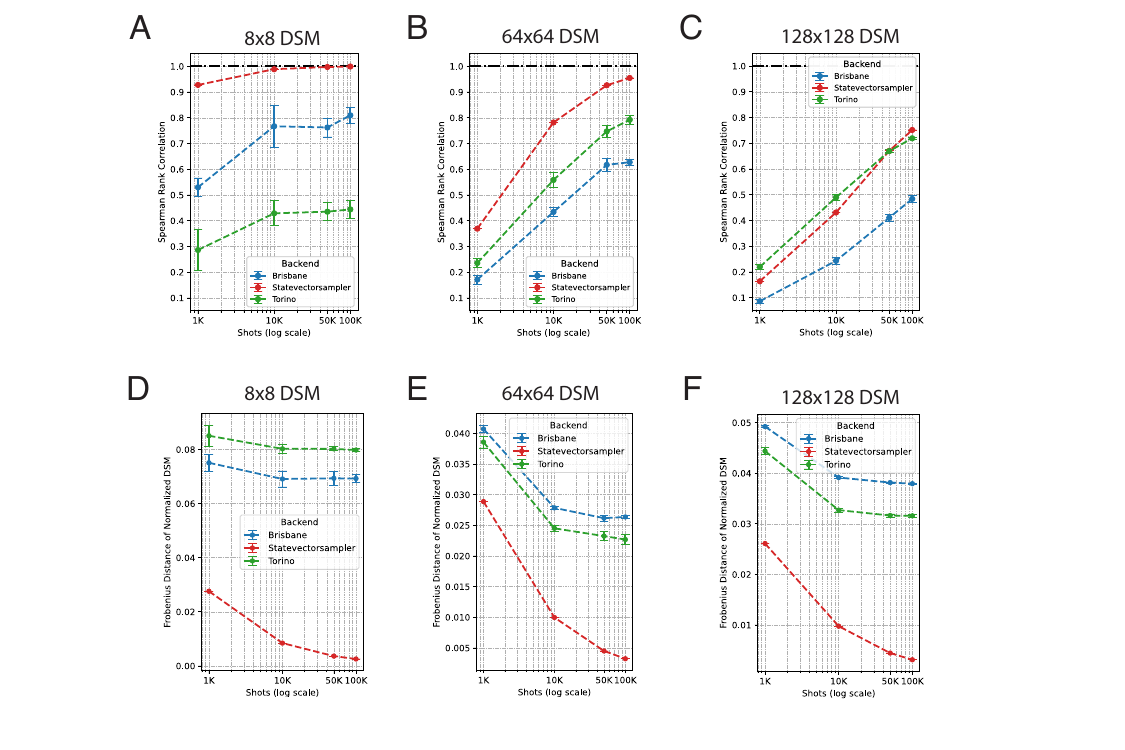}
    \caption{
    \textbf{Recovery of doubly stochastic matrices on quantum hardware.}
    DSMs computed on IBM Eagle R3 (\textit{ibm\_brisbane}) and IBM Heron R1 (\textit{ibm\_torino}) quantum processors were compared with exact classical solutions and noise-free statevector simulations.
    \textbf{a--c}, Spearman correlation between recovered and exact DSMs for matrix sizes $8 \times 8$, $64 \times 64$ and $128 \times 128$.
    \textbf{d--f}, Normalized Frobenius distance between recovered and exact DSMs for the corresponding matrix sizes.
    }
    \label{fig:dsm_hw}
\end{figure}

Across the tested matrix sizes, DSMs recovered from quantum hardware achieved Spearman correlations of at least 0.7 relative to the exact solution within 100{,}000 shots (Fig.~\ref{fig:dsm_hw}a--c). This level of agreement is notable given that worst-case theoretical shot requirements for accurate DSM recovery can be substantially larger~\cite{born2025quantum}. Spearman correlation was used because downstream attention behavior depends primarily on the relative ordering of attention weights rather than exact element-wise agreement. As expected, noise-free statevector simulations generally provided an upper bound on agreement with the exact DSM. At the largest tested matrix size (\(128 \times 128\); Fig.~\ref{fig:dsm_hw}c), however, hardware performance was comparable to and in some cases slightly exceeded the corresponding statevector result. This likely reflects the combined effects of finite-shot estimation and the rank-based nature of the metric, rather than a systematic hardware advantage. Overall, the gap between statevector and hardware results is consistent with the expected influence of sampling variability and hardware noise.

We also evaluated normalized Frobenius distance between recovered and exact DSMs to quantify element-wise deviations. The normalized distance did not increase substantially with matrix size across the tested settings (Fig.~\ref{fig:dsm_hw}d--f), suggesting that larger DSMs can be approximated within the tested shot budgets. This observation is encouraging because future applications of QDSM attention may require larger attention matrices than the $8 \times 8$ matrices used for classically simulated SEQUOIA training. However, the result should not be interpreted as evidence that arbitrarily large QDSMs can be generated without additional cost. Larger matrices may require deeper circuits, more qubits, more shots, improved compilation strategies or error-mitigation procedures, depending on the target accuracy and hardware topology.

Taken together, these experiments show that the DSM-generation primitive used by QDSM attention can be implemented on QPUs at matrix sizes relevant to transformer attention modules. 
They provide a hardware-level feasibility check for the quantum component of the model. In the context of computational pathology and genomics, this result is relevant because it supports a modular hybrid workflow: classical GPUs can perform feature extraction, transformer optimization and gene expression regression, while QPUs may be invoked selectively for structured attention normalization. Such a division of labor is consistent with near-term hybrid quantum-classical computing, where quantum subroutines are evaluated as targeted components within larger classical machine learning pipelines~\cite{seelam_reference_2026}.

\section{Discussion}
Histopathology is one of the most widely available sources of tumor information in clinical care, whereas transcriptomic profiling remains unevenly accessible because sequencing can be costly, tissue-consuming, and difficult to obtain in small biopsies, retrospective cohorts, and rare cancers. In this study, we evaluated whether a hybrid quantum-classical attention mechanism could improve molecular profiling from whole-slide histopathology within a practical digital pathology workflow. Across 29 TCGA cancer cohorts, the approach produced selective gains relative to standard attention, with the largest relative improvements observed in smaller and data-limited cancers. These gains were not uniform across the transcriptome, indicating that hybrid attention is best understood as a targeted modeling strategy for improving prediction of selected genes and pathways rather than as a universal replacement for conventional softmax attention. In the context of digital medicine, this distinction is important: the value of image-based molecular inference lies not only in maximizing aggregate performance, but in improving access to biologically and clinically relevant molecular information from routinely acquired pathology images when direct sequencing is unavailable.

A central finding of this study is that the hybrid attention mechanism redistributed predictive performance across the transcriptome rather than producing a broad transcriptome-wide increase in accuracy. In several cancer types, QDSM attention increased the number of genes whose expression could be more accurately inferred from histopathology and these genes were enriched for coherent biological processes. In other tumor contexts, standard softmax attention remained superior for subsets of targets. This behavior is consistent with the broader challenge of digital pathology-based molecular prediction: not all genes are equally coupled to tissue morphology, and the detectability of molecular states in H\&E images likely depends on tumor architecture, stromal context, lineage state, technical variation, and cohort composition. From a translational perspective, this means that the most plausible clinical role for the method is not transcriptome-wide replacement of sequencing, but selective molecular triage or augmentation of existing workflows, particularly for predefined pathways, signatures, or biomarker panels where image-derived information is most informative.

The prediction correlation improvement analysis supports this selective interpretation. Across most cancer types, the genome-wide median change in Pearson correlation was modest and frequently close to zero, indicating that the hybrid attention mechanism did not induce a large global shift in image-based molecular prediction. However, genes classified as improved under QDSM attention showed substantially larger positive effect sizes in selected cancers, including ACC, PCPG, and UVM, whereas worsened genes also showed non-trivial negative shifts in some cohorts. These findings suggest that the principal effect of the attention mechanism is to alter how image-derived signals are allocated across targets rather than to act as a uniform enhancer of performance. For digital medicine applications, these findings change how a model should be interpreted. Hybrid quantum-classical attention is not a universal replacement for softmax attention. Its value lies in focused molecular profiling, where the aim is recovering specific cancer-relevant gene programs, lineage markers, or clinically actionable signatures from histopathoogy images. In that setting, selective advantage can still be useful even if the average transcriptome-wide improvement is modest.

The gene-level modeling analysis provides a complementary view by separating components of QDSM-associated benefit that are partly predictable from baseline molecular properties from those that remain cancer- and pathway-specific. Using baseline predictivity, expression abundance, expression variance, and co-expression degree, the leave-one-cancer-out mixed-effects model showed moderate agreement between observed and predicted effect sizes, indicating that a meaningful fraction of the hybrid attention benefit can be anticipated from gene-level features alone. At the same time, the residual analysis revealed systematic cancer-specific deviations from these expectations. In particular, genes in UVM, ACC, PCPG, UCS, and KIRC tended to improve more than expected under QDSM attention even after accounting for baseline molecular predictors. Residual-based enrichment analyses further suggested that these unexplained gains were biologically structured rather than random, with developmental and morphogenetic programs among the pathways enriched for more-than-expected improvement, while selected DNA replication, repair, and adaptive immune pathways improved less than expected. These results reinforce the idea that hybrid attention is neither a uniformly beneficial nor wholly unpredictable intervention. Instead, it appears to produce partly predictable and partly context-dependent changes in image-based molecular inference that could be useful for prioritizing targets in future translational studies.

The ACC analyses provide a concrete example of where this selective effect may be practically useful. ACC is a rare and molecularly heterogeneous malignancy for which broad sequencing access is often limited and sample sizes are typically small. In this cohort, QDSM attention improved prediction for substantially more genes than it worsened, and the improved genes were enriched for biologically plausible processes including DNA damage and repair, transcriptional regulation, and ubiquitin-related signaling. These observations do not establish new biomarkers or causal mechanisms, but they suggest that structured attention may help recover selected molecular programs from routinely available histopathology in rare cancers where conventional model development is constrained by sample size. From a digital medicine perspective, this is one of the more promising use cases for the present approach: not generalized molecular replacement, but targeted image-based molecular profiling in cancers where both sequencing availability and training data are limited.

The ACC survival analysis further supports the interpretation that QDSM attention preferentially improves prediction of molecular features that are both biologically coherent and clinically relevant. All ten of the strongest survival-associated genes identified in ACC, including \textit{LYG1}, \textit{KIF18A}, \textit{LBR}, \textit{CENPL}, \textit{MAPK11}, \textit{NUMBL}, \textit{GPR89A}, \textit{SLC25A43}, \textit{ZBED4}, and \textit{HDAC4}, had hazard ratios greater than 1, indicating that higher expression was consistently associated with worse overall survival. This signature maps closely to aggressive ACC biology, especially mitotic regulation, chromosomal instability, and epigenetic control, and is consistent with prior evidence that poor-prognosis ACC is enriched for proliferative and DNA-damage–related programs. Among these genes, \textit{KIF18A} is particularly notable because prior work has already linked its overexpression to poor survival in ACC~\cite{liu_comprehensive_2023}, whereas several of the others are best considered emerging candidates supported by broader cancer biology rather than direct ACC-specific validation. The especially large hazard ratios observed for \textit{LYG1} and \textit{HDAC4} suggest strong adverse associations, although these likely reflect coordinated high-risk molecular states rather than fully independent effects. Taken together, these findings reinforce the view that hybrid attention may be most useful when it improves image-based inference of histopathologically encoded molecular programs that also carry prognostic information in rare cancers.

The external pancreatic cancer analysis offers a similarly cautious but informative translational signal. When models trained on TCGA-PAAD were transferred to an independent CPTAC pancreatic cohort, the hybrid attention mechanism improved prediction for a subset of genes that included metabolic regulators and lineage-associated transcriptional programs relevant to pancreatic cancer biology, while a larger number of genes showed reduced performance. This mixed result is realistic for digital pathology-based molecular profiling, where cohort composition, slide preparation, tissue processing, sequencing protocols, and tumor heterogeneity can all affect which molecular signals are recoverable from histology. The potential for quantum-classical hybrid attention is that some biologically relevant targets retained improvement outside the discovery cohort, suggesting that the  inductive bias can remain useful under domain shift. The practical implication is that hybrid attention should be evaluated as a targeted molecular profiling strategy, with prespecified genes or pathways and external validation, rather than as a universal solution for transcriptome-wide transfer.

The observed association between smaller cohort size and greater relative improvement is notable in the context of digital medicine because many clinically important settings involve limited data rather than large, idealized training cohorts. Rare cancers, uncommon molecular subtypes, understudied populations, and retrospective institutional collections all present constraints that can reduce the effectiveness of standard deep learning approaches. The doubly stochastic attention constraint may act as a structured regularizing bias by encouraging more balanced use of histopathology-derived tokens, thereby reducing overconcentration of attention and improving the use of limited information. This interpretation is consistent with prior work on doubly stochastic attention and with broader interest in machine learning strategies that are robust in limited-data settings~\cite{sander2022sinkformers,caro2022generalization}. The additional quantum motivation is that the QDSFormer construction can generate a diverse family of doubly stochastic matrices, offering an alternative structured attention prior whose utility may be greatest when data alone are insufficient to support reliable generalization~\cite{born2025quantum}. For the purposes of digital medicine, however, the critical point is not the abstract expressivity of the quantum construction, but whether it leads to better and more reliable molecular inference in use cases where clinical data are sparse.

The warm-start and hardware experiments suggest a possible practical path for hybrid quantum-classical digital pathology, while also underscoring how early such integration remains. Applying QDSM attention only during early training partially recapitulated the behavior of full-QDSM training, indicating that the structured attention prior can influence optimization even when applied for a limited portion of model training. Separately, the hardware experiments showed that the doubly stochastic matrix primitive underlying the attention mechanism can be recovered on IBM quantum processors for matrix sizes relevant to transformer attention modules. These results do not demonstrate end-to-end live quantum training of whole-slide pathology models, nor do they establish clinical utility from hardware-executed inference. Instead, they support a modular view of hybrid computing in which classical systems continue to perform feature extraction, regression, and most of model optimization, while quantum hardware may eventually contribute targeted structured operations within the broader AI pipeline. The practical implication is that the study is best interpreted as a translational AI investigation of a potentially useful structured attention mechanism.

Several limitations should guide interpretation. First, QDSM attention during SEQUOIA training was classically simulated and restricted to \(8 \times 8\) attention matrices for tractability, and larger matrices may capture richer token interactions as simulation and hardware capabilities improve. Second, we used a ResNet-50 image encoder rather than a stronger histopathology foundation model, so the absolute predictive performance reported here should not be interpreted as the upper bound of image-based molecular profiling. Third, gene-level improvements were used primarily for comparative model analysis and pathway prioritization and do not establish validated biomarkers without additional replication and endpoint-specific study design. Fourth, external transfer was evaluated in one tumor context only, and broader multi-institutional validation will be needed to assess robustness across cancer types, institutions, scanners, and slide-preparation workflows. Finally, the study did not include prospective deployment or workflow-based evaluation, so its implications for clinical implementation remain preliminary.

These limitations also define a focused translational and methodological agenda for future work. Rather than emphasizing transcriptome-wide reconstruction alone, future studies should evaluate hybrid attention on prespecified clinical signatures, targeted pathway scores, actionable molecular subtypes, or molecular triage tasks for which image-based estimates could inform downstream testing decisions. At the model-design level, our results also suggest that QDSM attention may be more useful as a complementary branch than as a universal replacement for softmax attention. A learned gated mixed attention framework could decide how much to rely on softmax versus QDSM attention for a given head, cancer type, pathway, or gene, preserving the stability of standard attention while selectively invoking the hybrid quantum-classical doubly stochastic branch where it adds value. This design would better match the pattern observed here, in which QDSM attention improved selected molecular targets but worsened others, and may be especially important under external transfer, where domain shift can change which molecular signals remain recoverable from histology. Future studies should therefore combine broader external validation with evaluation against modern pathology foundation models, calibration and uncertainty analysis, and careful characterization of failure modes under domain shift. In a realistic digital medicine workflow, the most plausible role of this approach is to complement, rather than replace, direct molecular testing.

Overall, our findings position hybrid quantum-classical attention as a selective strategy for molecular profiling from routine histopathology, particularly in cancers where sequencing access is limited and training cohorts are small. In ACC, the same target-specific behavior that improved molecular inference also captured genes associated with adverse overall survival, linking model benefit to prognostically relevant biology. This work provides a framework for integrating quantum-derived structured attention into digital pathology models and for evaluating when such hybrid approaches can support molecular triage, risk stratification, and precision oncology in data-limited settings.

\appendix
\section{Appendix}

\renewcommand{\thefigure}{S\arabic{figure}}
\renewcommand{\thetable}{S\arabic{table}}
\setcounter{figure}{0}
\setcounter{table}{0}

\begin{figure*}[!htb]
    \centering
    \includegraphics[width=1\linewidth]{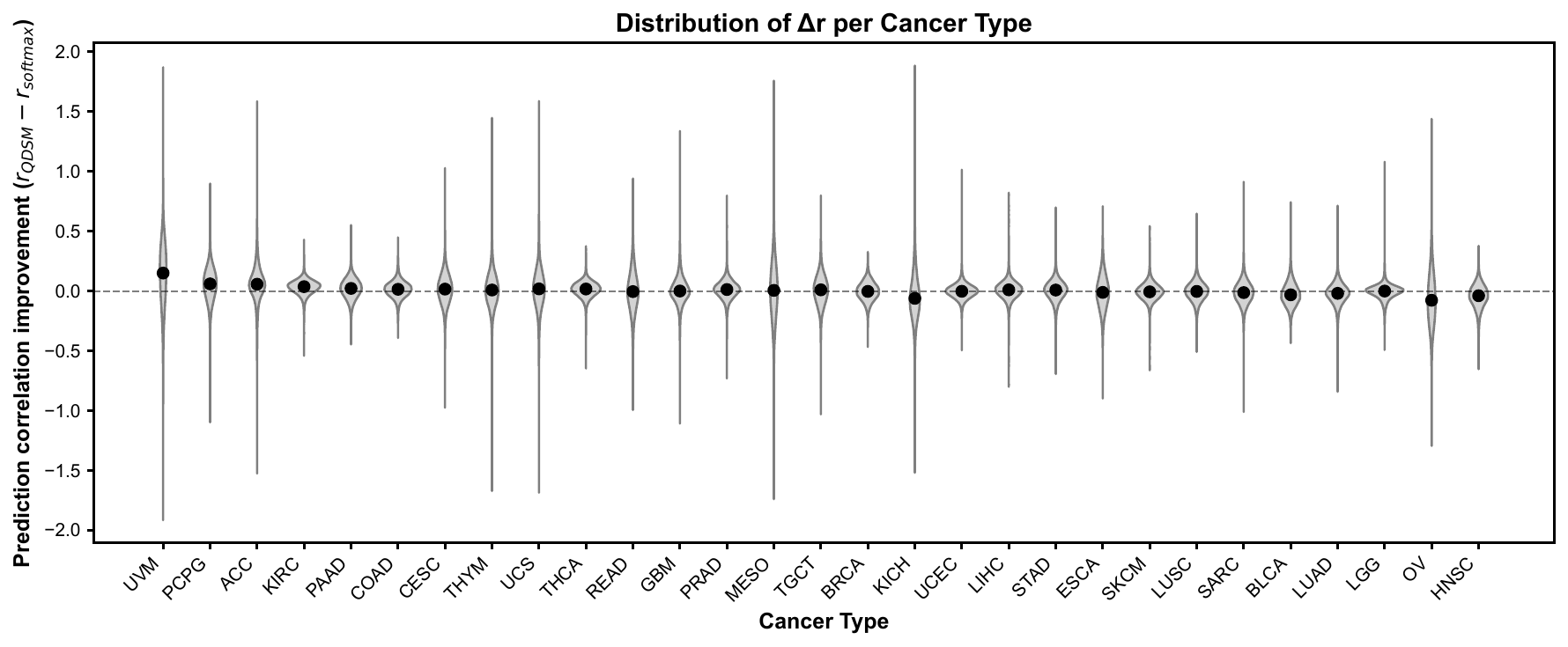}
    \caption{
    \textbf{Distribution of Pearson correlation improvement for QDSM attention across cancer types.}
    Violin plots show the distribution of gene-level changes in prediction performance after replacing softmax attention with QDSM attention in SEQUOIA. For each gene \(g\), the PCI was defined as
    $
    \Delta r_g = \mathrm{median}_{s=1,\ldots,10}\left(r^{\mathrm{QDSM}}_{g,s} - r^{\mathrm{softmax}}_{g,s}\right),
    $
    where \(r^{\mathrm{QDSM}}_{g,s}\) and \(r^{\mathrm{softmax}}_{g,s}\) denote the Pearson correlations between predicted and measured expression for gene \(g\) under QDSM and softmax attention, respectively, in train--test split \(s\). Positive \(\Delta r_g\) values indicate higher prediction accuracy under QDSM attention, whereas negative values indicate higher prediction accuracy under softmax attention. Cancer types are ordered from left to right in descending order of the \(\log_2\) ratio of significantly improved to significantly worsened genes. The dashed horizontal line denotes no change in performance (\(\Delta r_g = 0\)). Violin widths represent the relative density of genes at each \(\Delta r_g\) value, with internal markers showing the median and interquartile range. Confidence intervals for the median are not shown because they were narrow and fell within the plotted median markers for each cancer type.
    }
    \label{fig:effect_size}
\end{figure*}

\begin{table*}[htbp]
\centering
\caption{\textbf{PCI summary of QDSM attention across cancer types.}
For each cancer type, $\Delta r$ denotes the difference in Pearson correlation between QDSM and softmax attention, $\Delta r = r_{\mathrm{QDSM}} - r_{\mathrm{softmax}}$. Median values and confidence intervals are shown across genes. The columns ``Sig. up'' and ``Sig. down'' report the median $\Delta r$ among genes with significantly improved or worsened prediction, respectively. Percentages are reported as proportions. All values are rounded to two decimal places.}
\label{tab:deltar_effect_sizes}
\scriptsize
\begin{tabular}{lcccccc}
\toprule
\textbf{Cancer} & \textbf{Median $\Delta r$ (95\% CI)} & \textbf{Sig. up} & \textbf{Sig. down} & \textbf{IQR} & \textbf{$\Delta r > 0$} & \textbf{$\Delta r < 0$} \\
\midrule
ACC  &  0.06 [0.06, 0.06]   & 0.28 & -0.33 & 0.14 & 0.68 & 0.28 \\
BLCA & -0.03 [-0.03, -0.03] & 0.19 & -0.19 & 0.12 & 0.36 & 0.62 \\
BRCA & 0.00 [0.00, 0.00] & 0.12 & -0.13 & 0.10 & 0.48 & 0.50 \\
CESC &  0.02 [0.01, 0.02]   & 0.24 & -0.24 & 0.16 & 0.54 & 0.43 \\
COAD &  0.01 [0.01, 0.02]   & 0.13 & -0.14 & 0.08 & 0.58 & 0.39 \\
ESCA & -0.01 [-0.01, -0.01] & 0.28 & -0.27 & 0.19 & 0.46 & 0.52 \\
GBM  &  0.00 [0.00, 0.00]  & 0.22 & -0.21 & 0.12 & 0.49 & 0.48 \\
HNSC & -0.04 [-0.04, -0.04] & 0.14 & -0.19 & 0.12 & 0.33 & 0.65 \\
KICH & -0.06 [-0.06, -0.06] & 0.47 & -0.47 & 0.23 & 0.35 & 0.61 \\
KIRC &  0.04 [0.04, 0.04]   & 0.13 & -0.14 & 0.07 & 0.73 & 0.25 \\
LGG  &  0.00 [0.00, 0.00]  & 0.14 & -0.13 & 0.06 & 0.49 & 0.49 \\
LIHC &  0.01 [0.01, 0.01]   & 0.15 & -0.13 & 0.09 & 0.55 & 0.43 \\
LUAD & -0.02 [-0.02, -0.02] & 0.13 & -0.17 & 0.10 & 0.38 & 0.59 \\
LUSC & 0.00 [0.00, 0.00] & 0.14 & -0.16 & 0.10 & 0.48 & 0.50 \\
MESO &  0.01 [0.00, 0.01]   & 0.45 & -0.46 & 0.31 & 0.49 & 0.47 \\
OV   & -0.08 [-0.08, -0.07] & 0.41 & -0.40 & 0.30 & 0.36 & 0.61 \\
PAAD &  0.02 [0.02, 0.02]   & 0.15 & -0.15 & 0.11 & 0.60 & 0.38 \\
PCPG &  0.06 [0.06, 0.06]   & 0.25 & -0.24 & 0.18 & 0.66 & 0.32 \\
PRAD &  0.01 [0.01, 0.01]   & 0.14 & -0.16 & 0.09 & 0.57 & 0.41 \\
READ & 0.00 [-0.01, 0.00] & 0.28 & -0.31 & 0.21 & 0.47 & 0.50 \\
SARC & -0.01 [-0.01, -0.01] & 0.18 & -0.19 & 0.11 & 0.43 & 0.55 \\
SKCM & -0.01 [-0.01, 0.00] & 0.14 & -0.13 & 0.08 & 0.45 & 0.53 \\
STAD &  0.01 [0.01, 0.01]   & 0.16 & -0.18 & 0.10 & 0.53 & 0.45 \\
TGCT &  0.01 [0.01, 0.01]   & 0.23 & -0.24 & 0.16 & 0.52 & 0.45 \\
THCA &  0.02 [0.02, 0.02]   & 0.12 & -0.12 & 0.08 & 0.60 & 0.38 \\
THYM &  0.01 [0.01, 0.01]   & 0.28 & -0.26 & 0.17 & 0.51 & 0.46 \\
UCEC & 0.00 [0.00, 0.00] & 0.12 & -0.12 & 0.07 & 0.48 & 0.50 \\
UCS  &  0.02 [0.02, 0.02]   & 0.34 & -0.39 & 0.20 & 0.53 & 0.44 \\
UVM  &  0.15 [0.15, 0.16]   & 0.45 & -0.53 & 0.35 & 0.67 & 0.28 \\
\bottomrule
\end{tabular}
\end{table*}

\begin{figure*}[!t]
    \centering
    \includegraphics[width=0.9\textwidth]{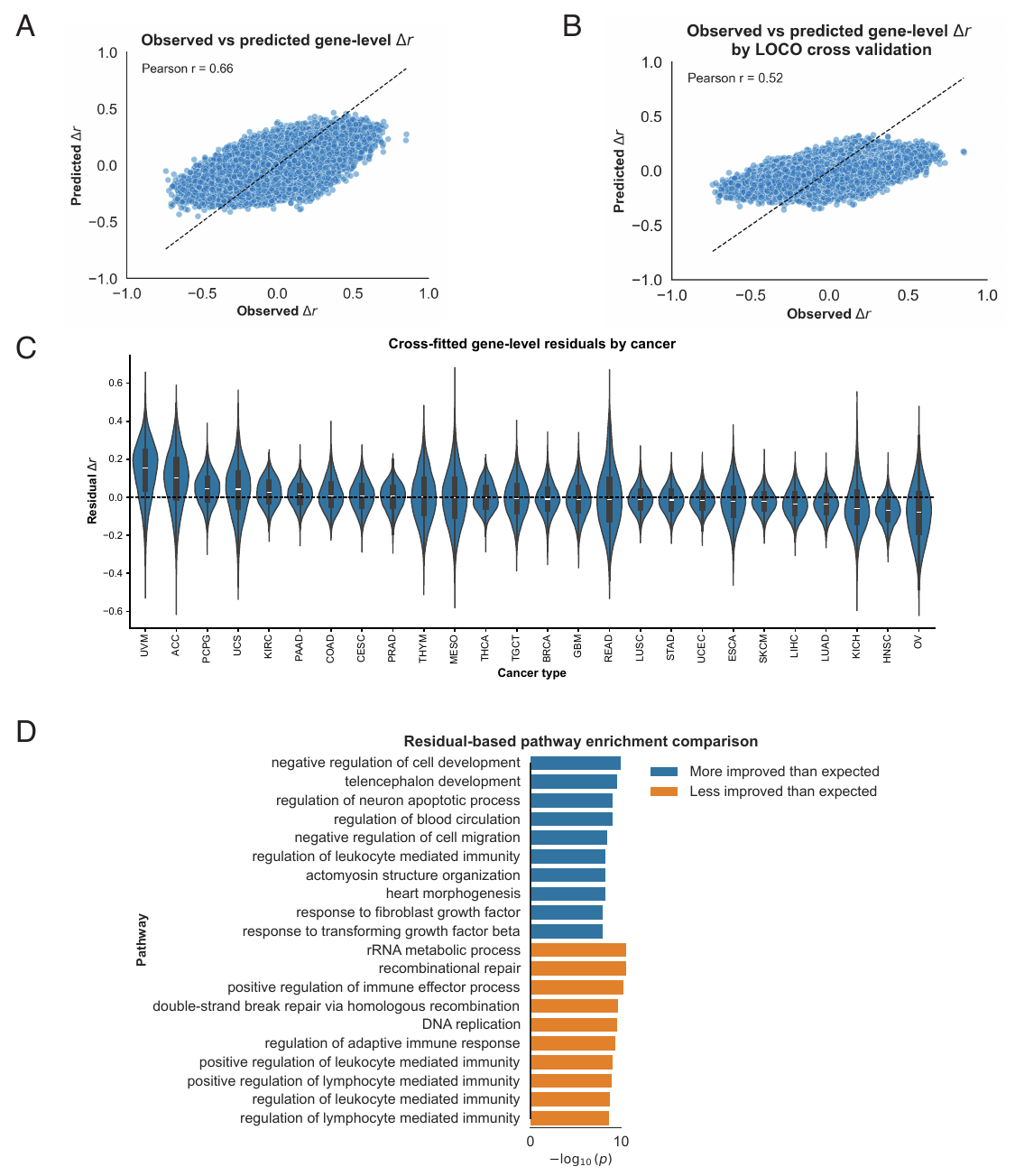}
    \caption{
    \textbf{Gene-level modeling of QDSM benefit and residual pathway structure.}
    \textbf{a}, Observed versus predicted gene-level PCI, defined as \(\Delta r = r_{\mathrm{QDSM}} - r_{\mathrm{softmax}}\), for the full mixed-effects model fit to all gene--cancer pairs. Each point represents one gene--cancer pair. Predictions include both the fixed effects and the cancer-specific random intercept.
    \textbf{b}, Observed versus predicted gene-level PCI for the leave-one-cancer-out (LOCO) model. Each point represents one gene--cancer pair in a held-out cancer cohort. Predictions are based on the fixed-effects component alone.
    \textbf{c}, Global distribution of residual \(\Delta r\) values from the LOCO model. Residuals were centered near zero and were used to identify genes that improved more or less than expected under QDSM attention after accounting for baseline predictivity, expression abundance, expression variance and co-expression degree.
    \textbf{d}, Pathway enrichment analysis of genes that improved more than expected or less than expected under QDSM attention, based on residuals from the LOCO model. The top 10 enriched pathways are shown for each direction. Positive-residual genes represent genes with larger observed \(\Delta r\) than predicted by the model, whereas negative-residual genes represent genes with smaller observed \(\Delta r\) than predicted. Pathway analysis was restricted to gene sets containing more than 10 and fewer than 300 genes.
    }
    \label{fig:mixed_model}
\end{figure*}

\clearpage

\bibliographystyle{unsrtnat}

\bibliography{ref}


\end{document}